\documentclass[iop]{emulateapj}

\slugcomment{To appear in The Astrophysical Journal}

\shorttitle{Milky Way Substructures in the NGVS Footprint}
\shortauthors{Lokhorst et al.}

\begin{document}


\title{The Next Generation Virgo Cluster Survey.  XIX. Tomography of Milky Way Substructures in the NGVS Footprint} 


   \author{Deborah Lokhorst\altaffilmark{1,10}, Else Starkenburg\altaffilmark{1,2,11}, Alan W. McConnachie\altaffilmark{3},
          Julio F. Navarro\altaffilmark{1,12},
          Laura Ferrarese\altaffilmark{3},
          Patrick C\^ot\'e\altaffilmark{3},
          Chengze Liu\altaffilmark{4,5},
          Eric W. Peng\altaffilmark{6,7},
          Stephen D.J. Gwyn\altaffilmark{3},
          Jean-Charles Cuillandre\altaffilmark{8},
          Puragra Guhathakurta\altaffilmark{9}
          }
          
\altaffiltext{1}{Department of Physics \& Astronomy, University of Victoria, Victoria, B.C., V8P 1A1 Canada }
\altaffiltext{2}{Leibniz-Institut f\"{u}r Astrophysik Potsdam, An der Sternwarte 16, 14482 Potsdam, Germany}
\altaffiltext{3}{National Research Council, Herzberg Astronomy \& Astrophysics, 5071 West Saanich Road, Victoria, B.C., V9E 2E7 Canada}
\altaffiltext{4}{Center for Astronomy \& Astrophysics, Department of Physics \& Astronomy, Shanghai Jiao Tong University, Shanghai 200240, China }
\altaffiltext{5}{Shanghai Key Lab for Particle Physics and Cosmology, Shanghai Jiao Tong University, Shanghai 200240, China}
\altaffiltext{6}{Department of Astronomy, Peking University, Beijing 100871, China}
\altaffiltext{7}{Kavli Institute for Astronomy \& Astrophysics, Peking University, Beijing 100871, China}
\altaffiltext{8}{CEA/IRFU/SAP, Laboratoire AIM Paris-Saclay, CNRS/INSU, Universit\'e Paris Diderot, Observatoire de Paris, PSL Research University, F-91191 Gif-sur-Yvette Cedex, France}   
\altaffiltext{9}{Department of Astronomy and Astrophysics, University of California Santa Cruz, 1156 High Street, Santa Cruz, CA 95064, USA}
\altaffiltext{10}{Email address: dml@uvic.ca}
\altaffiltext{11}{Canadian Institute for Advanced Research (CIFAR) Global Scholar}
\altaffiltext{12}{Canadian Institute for Advanced Research (CIFAR) Senior Fellow}

\begin{abstract}
The Next Generation Virgo Cluster Survey (NGVS) is a deep $u^*giz$ survey targeting the Virgo cluster of galaxies at 16.5~Mpc. 
This survey provides high-quality photometry over  an $\sim$ 100 deg$^2$ region straddling the constellations of Virgo and Coma 
Berenices. This sightline through the Milky Way is noteworthy in that it intersects two of the most prominent substructures in the Galactic
halo: the Virgo Over-Density (VOD) and Sagittarius stellar stream (close to its bifurcation point). In this paper, we use deep
$u^*gi$ imaging from the NGVS to perform tomography of the VOD and Sagittarius stream using main-sequence turnoff (MSTO) stars
as a halo tracer population. The VOD, whose centroid is known to lie at somewhat lower declinations
($\alpha \sim 190^\circ$, $\delta \sim -5^\circ$) than is covered by the NGVS, is nevertheless clearly detected in the NGVS 
footprint at distances between $\sim$ 8 and 25~kpc. By contrast, the Sagittarius stream is found to slice directly across the NGVS
field at distances between 25 and 40~kpc, with a density maximum at
$\simeq$~35~kpc. No evidence is found for new substructures 
beyond the Sagittarius stream, at least out to a distance of $\sim$~90 kpc --- the largest distance to which we can reliably trace the halo 
using MSTO stars. We find clear evidence for a distance gradient in the Sagittarius stream
across the $\sim 30$~deg of sky covered by the NGVS and its flanking fields. We compare our distance measurements along the
stream to those predicted by leading stream models.
\end{abstract}

\keywords{Galaxy: halo -- Galaxy: stellar content -- Galaxy: structure -- Local Group}

\section{Introduction}

Large-scale stellar mapping of the Milky Way by panoramic optical and infrared imaging surveys --- most notably the Sloan Digital Sky Survey (SDSS)
and the Two Micron All Sky Survey (2MASS) --- has revealed a rich and complex panoply of faint, remote substructures (e.g., \citealt{belo06,maje03,newb07,newb02,ibat01A,ibat01B,ibat02A,mart14}).
Indeed, the many stellar streams that have now been discovered  in the outer Galaxy provide {\it prima facie} evidence that
the assembly of the halo is an ongoing process. While some of these streams are clearly
still attached to their progenitor systems (e.g. the Sagittarius tidal
stream), others are seemingly isolated and left adrift away from
their ancestral homes \citep[e.g., the Orphan stream;][]{belo06,gril06}. Needless to say, the properties of these
stellar streams provide us with invaluable information on both the accretion history of the Milky Way and its gravitational potential.

The archetypal stellar stream --- that of the tidally disrupting Sagittarius dwarf galaxy
\citep[discovered by][]{ibat94,ibat95} --- has now been mapped
extensively across both the northern and southern hemispheres of the Galaxy
\citep[e.g.,][]{mate96,mate98,tott98,maje99,maje03,ivez00,yann00,dohm01,mart01,viva01,ruhl11, ibat01A, ibat02A, maje03,
newb02,bell03,newb03,belo06,yann09,corr10, nied10,kopo12,carl12b, jerj13,pila14,
belo14, kopo15, hyde15, huxo15}. 

As the quality and quantity of data for the stream has improved, a wide range of models for its formation have been
developed and refined
\citep[e.g.,][]{vela95,martinez01,helm01,ibat01B,helm04,law05,john05,fell06,law10,pena10,case12,deg13,ibat13,gibb14}. The
latest data now suggest that it is difficult to accommodate both the
kinematic and photometric measurements of the leading and trailing arm
within one single prolate, spherical or oblate Milky Way potential
\citep[although it is possible if the density profile of the halo is allowed
greater freedom,][]{ibat13}. 
From a modeling standpoint, the discovery of an apparent bifurcation in the stream
\citep{belo06,kopo12,slat13}, a recently discovered offset between the trailing arm and the predictions of
many models \citep{belo14,kopo15} and the possible existence of faint,
associated streams \citep{kopo13} have further complicated matters. 
At the present time, two widely used models for the Sagittarius tidal stream are those of
\citet[][hereafter LM10]{law10} and \citet[][hereafter Pen10]{pena10}. To lessen the tension between the leading and trailing arm data, these
models both use a triaxial --- though nearly prolate --- halo potential that has the 
awkward property of not being aligned with the plane of the Milky Way disk. The Pen10 model
was specifically focused on modelling the stream bifurcation, which they attribute to a
moderate amount of rotation in the stream progenitor. However, no evidence for such rotation has been
found in the core of the Sagittarius dwarf galaxy \citep{pena11}. 

More recently, \citet[][hereafter VC13]{vera13} have calculated the orbit of the Sagittarius dwarf galaxy in a Galactic
potential that includes both a varying shape with radius --- as is seen in cosmological simulations \citep{vera11} --- and  
the influence of the Large Magellanic Cloud (LMC).  While not a full N-body simulation, a comparison of their 
test particle orbit to the latest data showed that this new potential can alleviate some of the earlier tensions. It also 
showed that the inclusion of the LMC potential can significantly change the orbit of the Sagittarius and its tidal stream. 
In addition, \citet{gome15} have shown that in such a scenario the Milky Way's response to the orbit of the 
LMC must also be taken into account. As we shall argue below, new simulations of the Sagittarius stream including all 
these effects are now needed, while deeper imaging at selected positions along the stream, including direct
distance measurements, are needed to constrain the various model parameters. 

   \begin{figure*}
   \centering
   \includegraphics[scale=0.76]{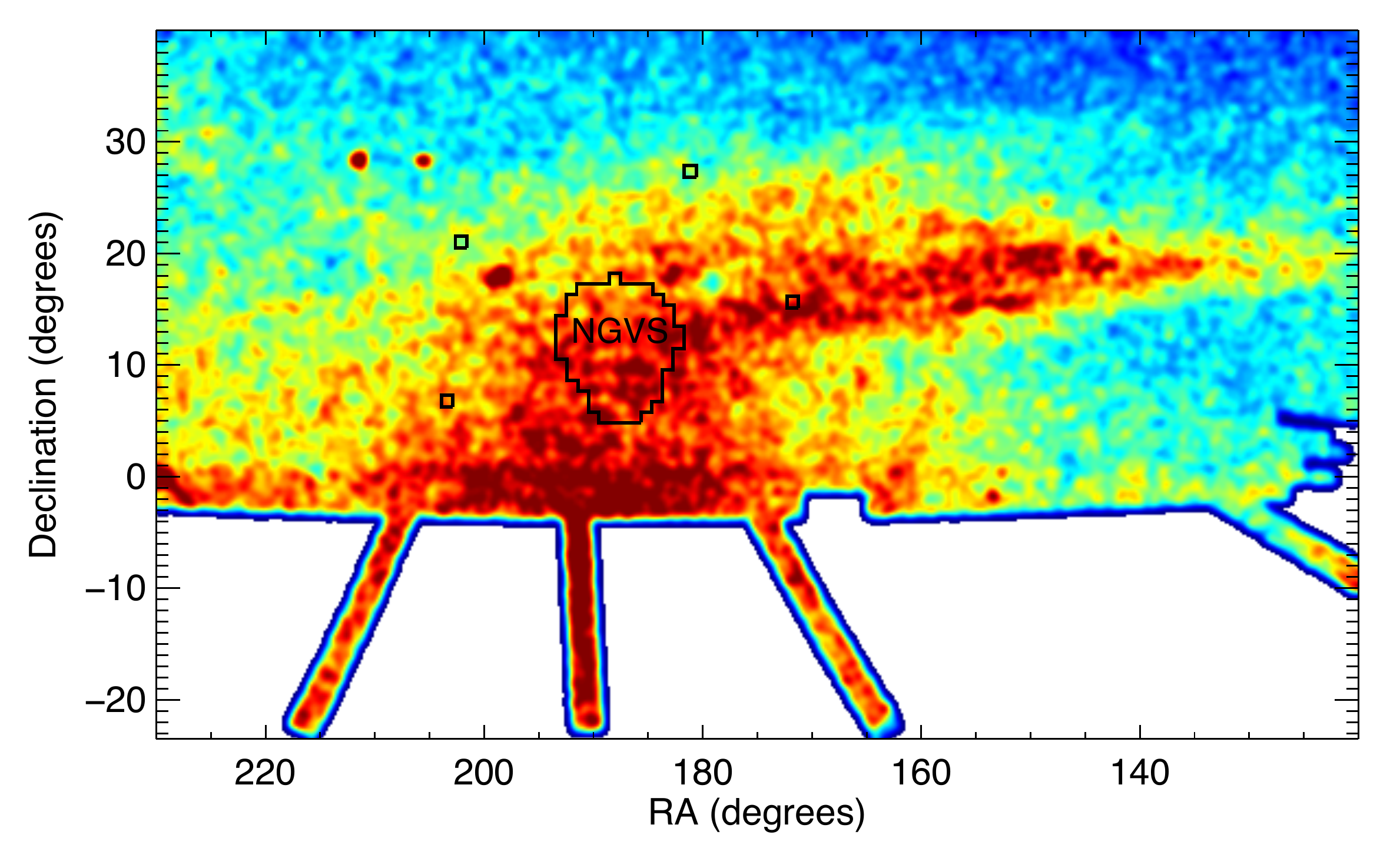}
   \caption{A stellar density map made using SDSS stars selected to have colors of 0.2 $<$ ($g_{\tiny\textnormal{SDSS}}-r_{\tiny\textnormal{SDSS}}$) $<$ 0.3 
   		and ($u_{\tiny\textnormal{SDSS}}-g_{\tiny\textnormal{SDSS}}$) $>$ 0.4, and magnitudes of 20.4 $<$ $g_{\tiny\textnormal{SDSS}}$ $<$ 21.4.  This 
		selection showcases the various large-scale Milky Way substructures in the northern hemisphere that happen to fall along the line of sight to the 
		constellation of Virgo. The location of the 100 deg$^2$ NGVS survey footprint and its four background regions are shown.  The NGVS
		lands squarely on the northern tidal stream of the Sagittarius dwarf galaxy, with the two lower background fields also falling on the tidal stream.
		}
              \label{FigBelNGVSFOS}%
    \end{figure*}
%


A second prominent substructure in this region of the sky, but one lying at a closer distance, is the Virgo Stellar Stream or 
Virgo Over-Density (hereafter VOD). The VOD is now recognized to span more than 1000 square degrees \citep{juri08,jerj13}
and contain several dense clumps embedded within it \citep{viva01,newb02,viva03,viva06,newb07,kell08,kell09,kell10}. 
In addition to the large-scale excess of halo stars that was originally used to identify the VOD, several distinct, kinematically-grouped 
substructures have now been found in this region \citep[][]{duff06,newb07,viva08,prio09,star09,brin10,duff10,case12,duff14}. 
Although most of these kinematical structures show high positive galactocentric velocities, their spread in velocity
can be more than 100 km s$^{-1}$. The nature of these over-densities --- and their 
relationship to the VOD itself and to each other --- has yet to be conclusively established. None of them seems
obviously associated with the Sagittarius leading arm, the bifurcation feature, or any of the predicted trailing arm 
features or older wraps in this region. It has been suggested that many, and perhaps all, of these substructures could be
explained as the remnants of a now disrupted satellite galaxy (e.g., \citealt{carl12} and references therein) with stellar 
proper motion measurements for some of the substructures seeming to point to an orbit that has just passed 
pericenter \citep{case06, carl12}. Whatever the true explanation, it is clear that the sightline towards the constellation 
of Virgo is a complex and intriguing ``crossroad" in the Milky Way --- and one for which a consensus has yet to emerge.

By a lucky coincidence, this region of the sky is also home to the rich cluster of galaxies nearest to the Milky Way: the Virgo
cluster, at a distance of 16.5~Mpc \citep{mei07,blakeslee09}. The cluster has been the subject of numerous past imaging studies 
(see, e.g., \citealt{richter85, binggeli85, cote04} 
and references therein), and is the target of a new, deep, panoramic, multi-band survey with the Canada-France-Hawaii
Telescope (CFHT): the Next Generation Virgo Cluster (NGVS; \citealt{ferr12}).
Although the primary goal of the NGVS is a census and characterization of baryonic substructure 
{\it within}  the Virgo cluster, the survey is also ideally suited for studying the structure of the Milky Way halo along this direction. 
For instance, in their NGVS study of Virgo's globular cluster populations, \citep{durr14} clearly identified the Sagittarius 
stellar stream as a prominent foreground feature (see their Figure~5). Our study thus builds upon the tradition of 
using wide-field photometric data sets acquired for background galaxies or clusters to study the intervening halo: 
those of, e.g., \citet{mart14}, who mapped the highly structured Galactic foreground 
within the deep Pan-Andromeda Archaeological Survey \citep{mcco09}, and \citet{pila14}, who 
determined distances along the Sagittarius stream and various other substructures from pencil-beam 
survey data targeting galaxy clusters. 

In this work, we focus on the two main stellar over-densities in, and around,
the NGVS region: the Sagittarius stellar stream and the VOD. Using
old, metal-poor stars located near the main sequence turn-off (MSTO) region, we are able to measure
accurate distances to these over-densities and explore their three dimensional structure within the Galaxy.
These measurements allow us not only to probe the geometry of these substructures but to test the predictions
of numerical models that hinge on the assumed properties of their progenitors and the Galactic halo potential.

This is the first NGVS paper to focus specifically on the Milky Way foreground star population. Other papers in 
this series have examined the distribution of globular clusters within the Virgo cluster \citep{durr14}, 
the properties of star clusters, UCDs and galaxies in the cluster core \citep{zhu14, zhang15,liu15},
the internal dynamics of low-mass galaxies \citep[Toloba et al., submitted]{guerou15},
abundance matching of low-mass galaxies in the cluster core (Grossauer et~al. 2015, in press), 
interactions within possible infalling galaxy groups \citep{paudel13},
optical-IR source classification methods \citep{munoz14}, 
a new member of the inner Oort cloud \citep{chen13} and
a catalogue of photometric redshifts for background sources \citep{raichoor14}.

The paper is organized as follows. We begin with a short introduction to the NGVS in \S2.  In \S3, we give an overview of the 
main substructures visible in the very deep NGVS data, including the VOD and the Sagittarius tidal stream.  In \S4, we determine 
accurate distances to the Sagittarius tidal stream using the NGVS data and compare those distances to the predictions of three 
leading stream models.  We summarize and conclude in \S5.

 \section{Observations and Data}

The NGVS is a multi-band, panoramic imaging survey of the Virgo cluster carried out with the MegaCam instrument mounted at prime focus on
the 3.6m Canada-France-Hawaii Telescope (CFHT). Full details on the survey design, reduction procedures, data products and 
science goals have been presented in \citet{ferr12}. 

In brief, the survey covers an area of 104 deg$^2$ inscribed within the virial radii 
of Virgo's two main subclusters: the A subcluster to the north, centered on Virgo's cD M87, and the B subcluster to the 
south, centred on Virgo's optically brightest galaxy, M49. Observations were carried out in four optical bands ---
$u^*$, $g$, $i$, and $z$ --- in the MegaCam filter system\footnote{Note that the MegaCam bands have  similar names to those 
of the SDSS, but the filters do not cover exactly the same wavelength range nor do they have the same response 
function: see {\tt http://www.cadc.hia.nrc.gc.ca/en/megapipe/docs/filt.html}.} to 10$\sigma$ point-source depths of 
$u^*= 24.8$, $g = 25.9$, $i=25.1$ and $z=23.3$ AB mag. The survey has sub-arcsec 
seeing in all bands, with a median seeing of 0\farcs54 in the $i-$band.  The survey also includes partial coverage of the NGVS
footprint in the $r$ band (see, e.g., \citealt{raichoor14}).

In Figure~\ref{FigBelNGVSFOS}, we show the location of the NGVS footprint within
the large-scale Milky Way stellar halo in the Northern hemisphere, as seen by SDSS. 
In this figure, SDSS DR7 was used to create a 
density map of MSTO stars selected by the following color criteria: 
\begin{equation}
\begin{array}{rrcll}
0.2 & < & (g_{\rm SDSS} - r_{\rm SDSS}) & < 0.3 \\
      &     & (u_{\rm SDSS} - g_{\rm SDSS}) & > 0.4. \\
\end{array}
\end{equation}

A magnitude cut of 20.4 $<$ $g_{\tiny\textnormal{SDSS}}$ $<$ 21.4 was also applied in order to select SDSS stars having approximate distances of
15~$\lesssim$~$d_{\odot}$~$\lesssim$~35 kpc. In this distance range, the two over-densities that we focus on in this paper ---
namely, the VOD and the brighter branch of the Sagittarius tidal
stream bifurcation --- are unmistakable. The Sagittarius tidal tail is running diagonally from 
$(\alpha,\delta) \sim$ (120$^\circ$, 20$^\circ$) to (200$^\circ$,
10$^\circ$). The fainter bifurcation feature is seen above the main
stream. By contrast, the VOD appears as
an immense over-density which peaks at $(\alpha,\delta)\sim$ (190$^\circ$, --5$^\circ$). Although its center lies well south of our fields, the VOD
also extends northward into the NGVS footprint. Note that the NGVS also includes four
outlying background fields (see the four squares in Figure~\ref{FigBelNGVSFOS}
and \citealt{ferr12}) located $\sim$
16$^\circ$ from the center of the footprint. Among the four NGVS background fields, two lie well off the main portion of the Sagittarius tidal
stream, but two others fall squarely along the stream. As we show below, these two fields turn out to provide important constraints on  the 
overall geometry of the stream.

Our analysis of these substructures obviously requires a catalog of stellar sources selected from the NGVS. Within the NGVS footprint, a catalog optimized for 
compact and unresolved sources was created as described in \citet{liu15} and summarized below. Note that the NGVS has 
implemented several data processing pipelines, each optimized for a specific goal; for the purpose of this work, a source catalog was generated using 
NGVS stacks processed through the {\it MegaPipe} \citep{gwyn08} pipeline which adopts a global estimation of the sky background. For 
compact and unresolved sources, these stacks provide the highest photometric accuracy (see \citealt{ferr12} for details). Photometry was then 
performed using SExtractor \citep{bertin96} in dual-image mode with the $g-$band --- the deepest of the NGVS bands --- as the detection image. 
Aperture magnitudes were measured in apertures of diameter  3, 4, 5, 6, 7, 8, 16 and 32 pixels (each pixel corresponding to 0\farcs187) and then corrected to 
an infinite aperture. This last step was performed by applying an aperture correction measured by matching the MegaCam 16-pixel aperture magnitudes 
to SDSS PSF magnitudes (transposed to the MegaCam photometric system as given in equation 4 of Ferrarese et~al. 2012) for a number of bright but unsaturated stars. 
All other aperture magnitudes (within 3, 4, 5, 6, 7, and 8 pixels) were then corrected to a 16-pixel aperture. Because of PSF variations from field to field, 
aperture corrections were calculated separately for each field and, of course, for each filter.   

Point sources were then selected as having $-0.1 \leq (m_4-m_8) \leq  +0.15$~mag, where $m_4$ and $m_8$ are the corrected
aperture magnitudes measured within 4 and 8 pixels, respectively. The
smaller of these two apertures is well matched to the average seeing
of the MegaCam images given the pixel scale. The PSF is
sufficiently uniform over the images that all point sources
cluster tightly in $(m_4-m_8)$ (the same is generally true for the difference in
any two aperture magnitudes). Although point-source selections were carried out separately in each of the 
$u^*$, $g$ and $i$ bandpasses, the overall results are not affected by the choice of bandpass (or a combination of them). Therefore, for the 
remainder of this work, we use the catalog of stellar sources derived in the $g$ band. As a final step in the catalog preparation, the photometry 
for each object was de-reddened following \citet{schl98}. Note that the reddening along this line of sight is quite low (see, e.g., \citealt{boissier15}), with M87,
which marks the center of the Virgo cluster, having $E(B-V) = 0.022$~mag.

\begin{figure}
   \centering
   \includegraphics[width=9.46cm]{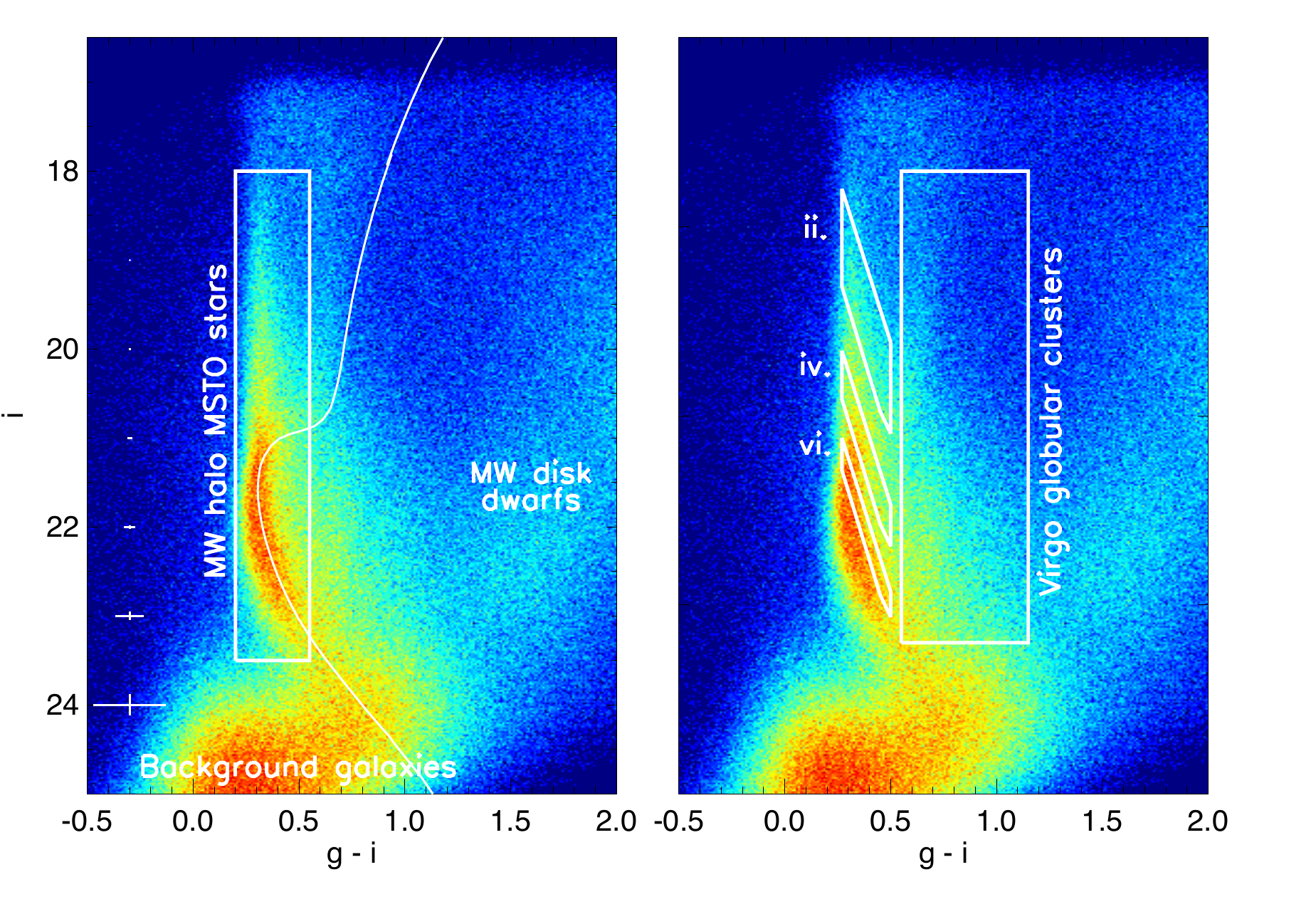}
   \caption{Color-magnitude diagrams (CMDs) for compact sources in the NGVS.  
   		\emph{(Left Panel)}  In this CMD, the location of halo main sequence turnoff (MSTO) stars, 
		disk dwarf stars and compact background galaxies are labelled.  
		An over-dense main sequence with a turnoff at $i \sim 21$ is clearly visible, 
		corresponding to the tidal stream of the Sagittarius dwarf galaxy.
	     	The best-fit theoretical isochrone for the tidal stream is shown as a reference. Typical
		photometric error bars are shown along the left side of the diagram.
		\emph{(Right panel)} In this CMD, three representative polygons used to select main 
		sequence turnoff stars are shown, corresponding to distances (from bright to faint) 
		of $\sim 10, 20$ and $30$~kpc.  Each main sequence selection region corresponds to the 
		similarly labelled density map in Figure \ref{FigMaps}.  The main locus of red globular clusters 
		in the Virgo cluster is also
		indicated, as the bluest globular clusters have the same ($g - i$) color as halo main sequence stars. 
             	 }
         \label{FigCMD}
   \end{figure}

Figure \ref{FigCMD} shows the (Hess) colour-magnitude diagram (CMD) of all NGVS stellar (point-like) sources identified in this way. 
The average photometric error in $i$ and $(g - i)$ as a function of $i$-band
magnitude is shown on the left-hand side of the figure.
This figure illustrates the location of Milky Way stars along this line of sight, which is
dominated by faint MS halo stars, plus some very red disk dwarfs in the solar neighbourhood.
At the faintest magnitudes, one sees residual contamination from compact,
background galaxies. At brighter magnitudes, still another source of contamination is discernible: globular clusters
belonging to the Virgo cluster which appear as nearly point-like at this
distance \citep{jordan05, durr14}. We will discuss our treatment of these various contaminants below.

%
%

%
   \begin{figure}
   \centering
   \includegraphics[width=9.4cm]{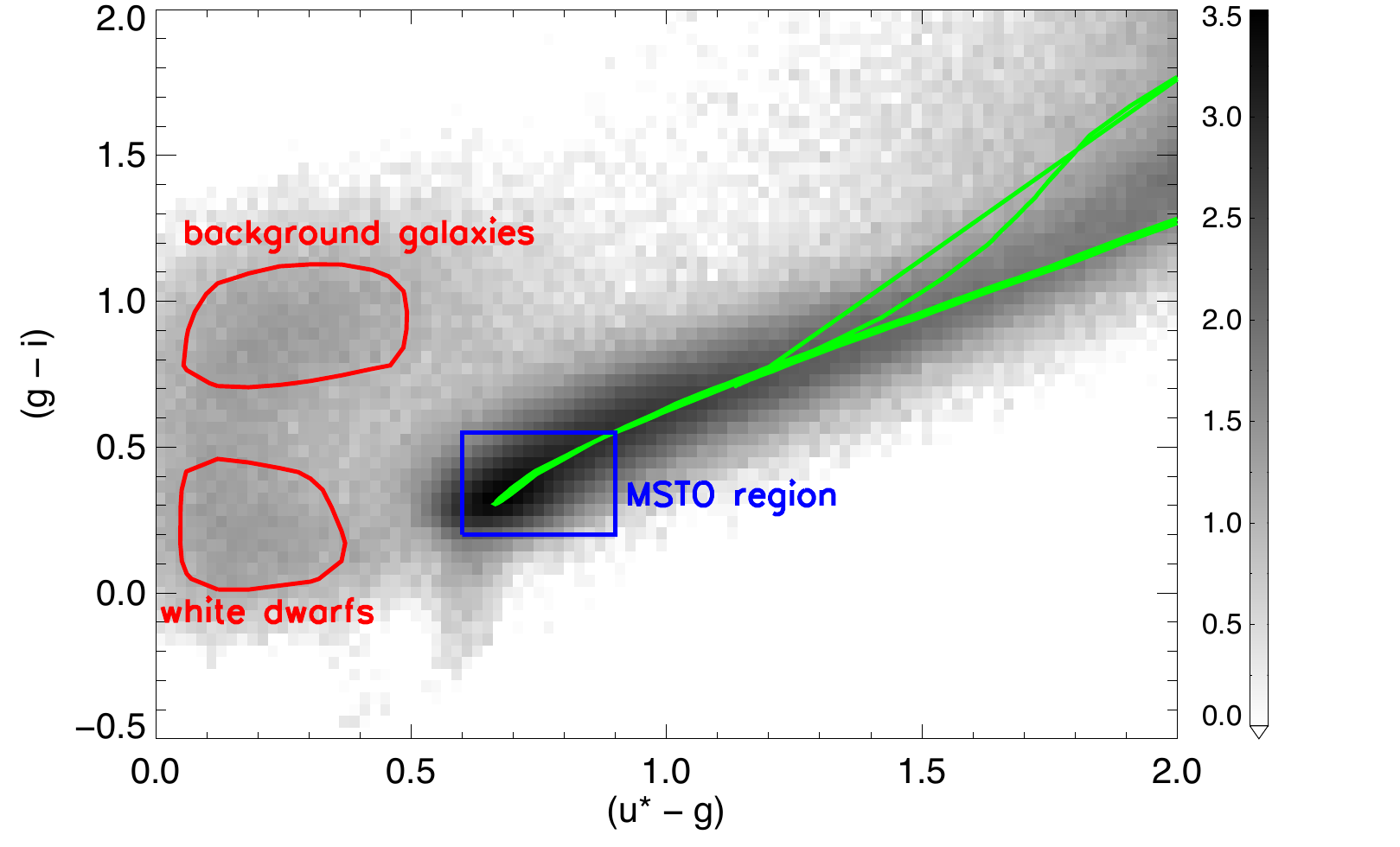}
   \caption{A colour-colour diagram of the NGVS stars
    		scaled with density inside bins of 0.1 mag.  The MSTO color
      		selection is shown in blue while the green curve shows a PARSEC isochrone with an
		age of 9 Gyr and a metallicity of [M/H] = --0.7 dex.  
		The main region occupied by background galaxies is also shown.
		Red contours surround their location and the locus of white dwarfs in the color-color plane.
		A ($u^{*}-g$) MSTO color cut was used
		to separate white dwarfs stars from the MSTO stars, which overlap in ($g-i$) color.
             	 }
         \label{FigColourColour}
   \end{figure}

%

   \begin{figure}
   \centering
   \includegraphics[width=8cm]{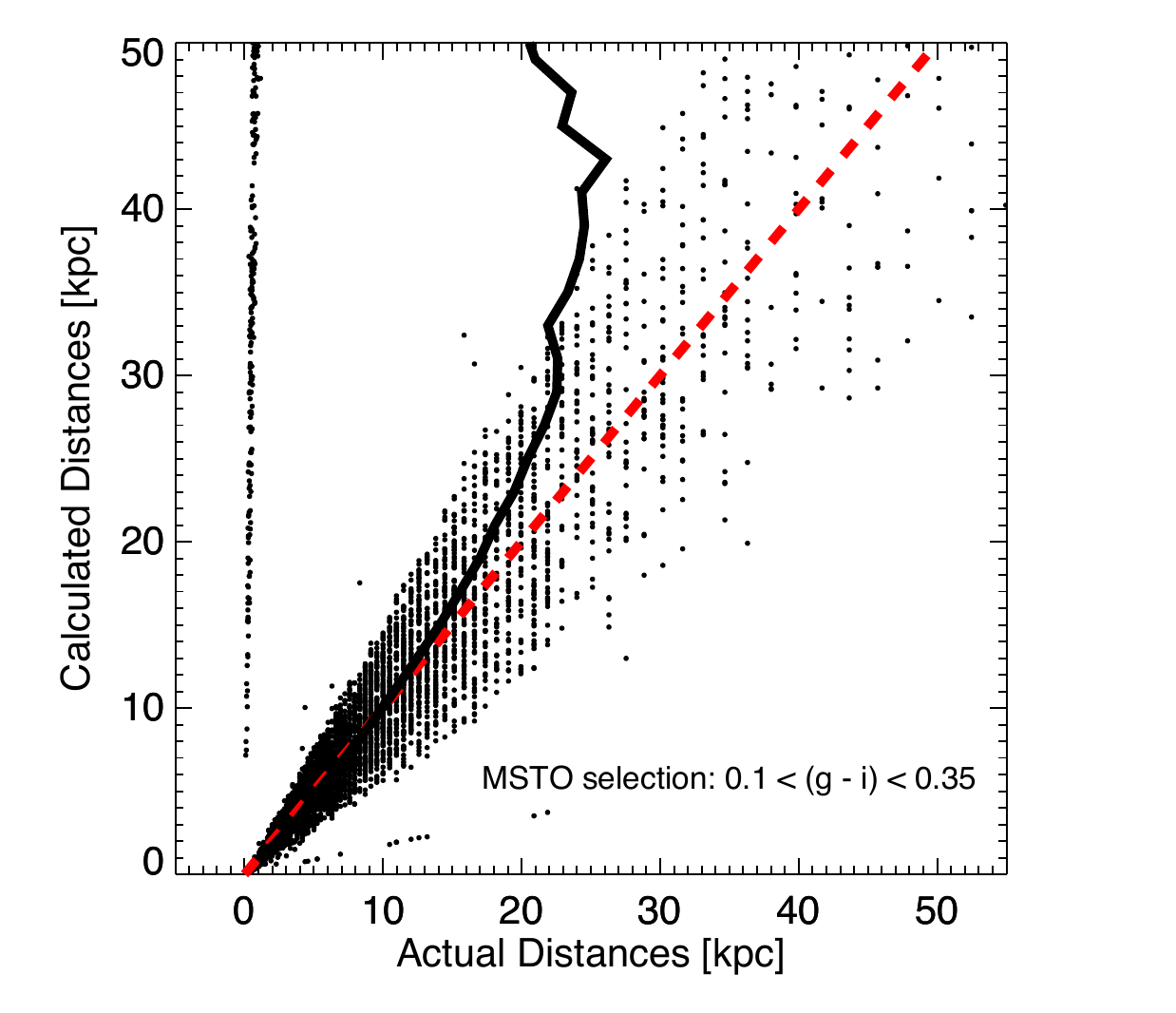}\\
   \includegraphics[width=8cm]{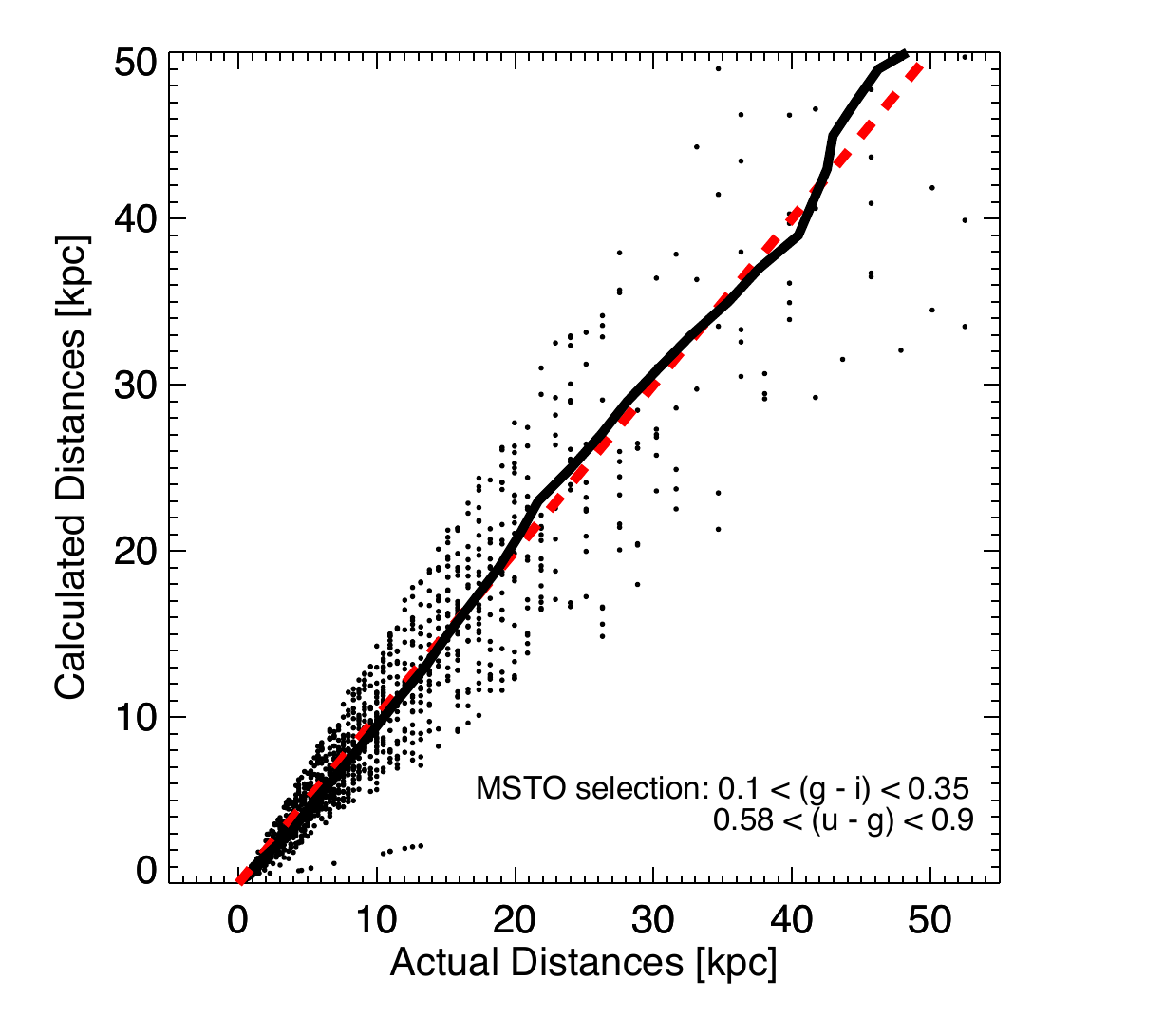}
    \caption{Distances derived from a mock catalog of stars taken from the
      TRILEGAL model of the Galaxy in the direction of NGVS (using
      the NGVS color band system and magnitude range). The distances were derived
      assuming a MSTO magnitude of $i$ = 4 ---  the
      approximate MSTO magnitude of halo stars (age $\sim$ 11 Gyrs, [Fe/H] $\sim$ -1.0) --- and plotted against the
      input distances from the mock catalog. In the upper panel,
      a ($g-i$) cut has been applied to the mock sample; in the lower
      panel, an additional ($u^{*}-g$) cut has also been applied. 
      The solid black line is the average calculated distance vs. actual distance, displaying 
      the trend's agreement with the dashed red one-to-one line.   It is clear from a
      comparison of the running averages that, without a ($u^{*}-g$) color cut to
      weed out white dwarfs in the sample, the derived distance profile
      will be highly skewed. }
         \label{WDdistcomp}
   \end{figure}
%


   \begin{figure*}
   \centering
   \includegraphics[scale=1.]{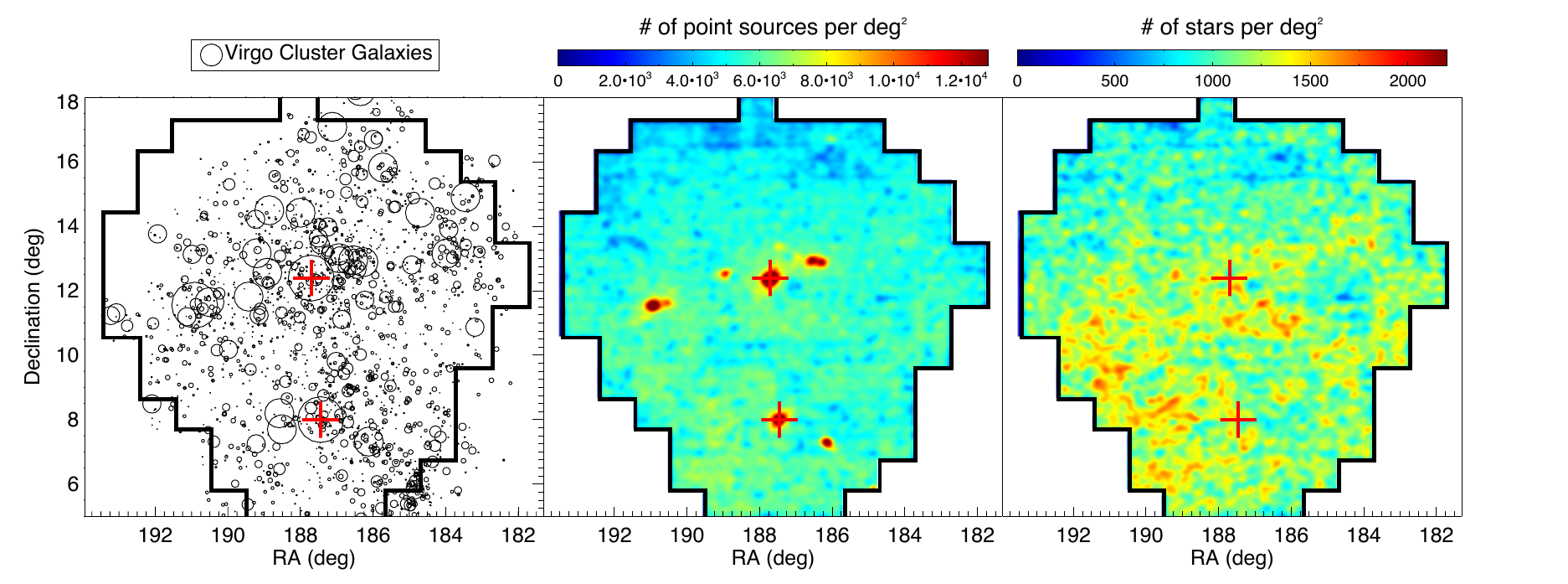}
    \caption{\emph{(Left panel)} The location of the Virgo cluster galaxies inside the NGVS footprint. The symbol size has
    		been scaled by galaxy $B$-band luminosity.  The red crosses mark the locations of M49 and 
		M87, which are located near the centers of subclusters A and B.  
		\emph{(Middle panel)} A density map of all NGVS compact sources with $i$ $<$ 23.5. M49 and 
		M87 are again marked by red crosses.  The obvious density enhancements at the positions of the brightest Virgo galaxies
		are due to contamination of the stellar catalogue by Virgo globular clusters.
		\emph{(Right panel)} A density map of main sequence stars 
		from the NGVS, selected using the color cuts given in
                Equations~2 and 3.  In this map, there are no longer any significant 
		over-densities at the locations of M49, M87 (red crosses) or any other Virgo galaxies: 
		i.e., the MSTO selection is highly effective in removing globular clusters from the stellar sample.
             	 }
         \label{FigGalaxies}
   \end{figure*}
%

   \begin{figure*}
   \centering
   \includegraphics[scale=1.3]{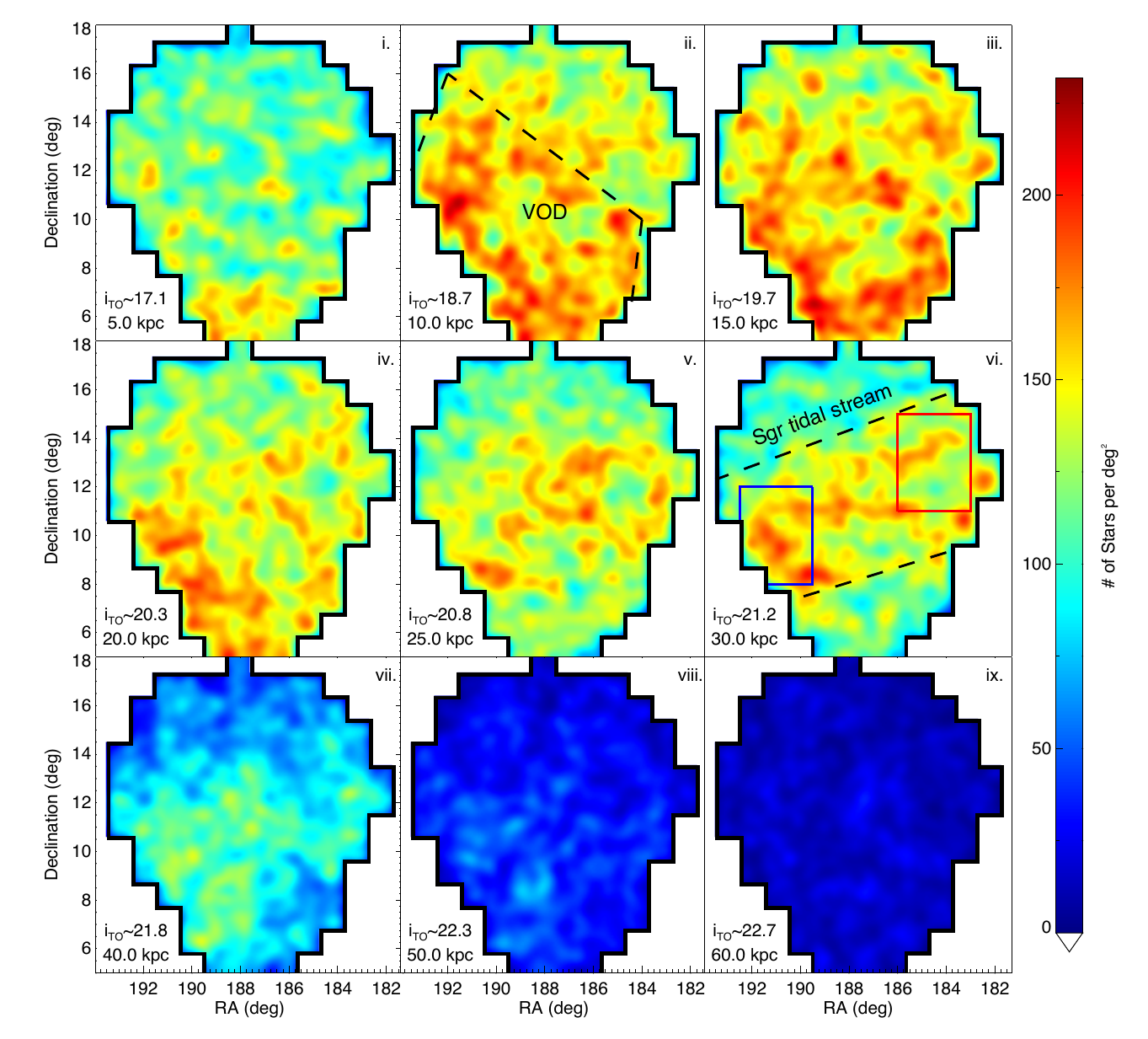}
   \caption{Density maps of MSTO stars in the NGVS footprint.  The nine panels step outwards in heliocentric distance through the 
   		Milky Way halo, displaying MSTO stars selected within boxes tailored around the 
		main sequence of a PARSEC isochrone.  A color cut in ($u - g$) has been used to eliminate white dwarfs from the 
		MSTO star sample (see Figure \ref{FigColourColour}).
		The boxes used to select stars in {\it panels ii}, {\it iv}, and {\it vi} are shown in Fig \ref{FigCMD}.  
		The width of the boxes in magnitude are scaled with distance such that the approximate range 
		in distance within each box is 5 kpc.  The mean distance and MSTO magnitude of the stars
		within each density map are labelled in the lower left of each panel. 
		The stars have been binned within 3\arcmin~$\times$ 3\arcmin ~pixels and smoothed with a Gaussian filter having FWHM = 12\arcmin.
		As the stars increase in distance from the disk, the feature referred to as the Virgo Over-Density (VOD) 
		appears at $d_{\odot}\sim$~8 kpc ({\it panel ii}).  The VOD increases in density and spatial extent at slightly larger 
		distances ($d_{\odot}\sim$~15 kpc), and then diminishes beginning around $d_{\odot}\sim$~20 kpc.  
		In {\it panel vi}, we see the Sagittarius tidal stream clearly flowing from low to high declination as RA decreases.
		In order to investigate the change in the Sagittarius tidal stream properties across the NGVS, the two spatial boxes shown in {\it panel vi}
		are used to determine an overall trend in stream distance. At distances well beyond the stream, we see a smooth spheroid 
		halo with no other significant substructure ({\it panels viii} and {\it ix}).}
              \label{FigMaps}%
    \end{figure*}

\section{Exploring Milky Way halo Substructure} \label{sec:explMWsubstr}

\subsection{Selecting a Tracer Population}

In this work, we use main sequence (MS) stars located near the MSTO as our stellar tracers.  MSTO stars are not as accurate distance indicators 
as some other stellar tracers (e.g., blue horizontal branch = BHB stars), but
they are more numerous than other stellar types and, as such, are attractive targets for mapping large-scale structure. 

To map halo features using  MS stars, we begin by constructing a polygon in the CMD centered on the MS of a 
particular theoretical isochrone (an ``MS box") in order to select stars located near the turn-off region.  Each  box is
centered on an isochrone that we henceforth adopt as the fiducial for this analysis: a PARSEC isochrone having 
an age of 9 Gyr and a metallicity of [M/H] = --0.695 dex \citep{bres12}, which has MSTO $i$-band absolute magnitude = 3.8. These values have been chosen to match
the Sagittarius stream properties in this region of the sky \citep{Chou07}. In the left panel of Figure \ref{FigCMD}, the 
PARSEC isochrone is overplotted on top of the most visible turn-off feature that we see within the NGVS CMD: 
that of the Sagittarius stream. Grafted onto this isochrone, our MS boxes are wide enough to accommodate MS 
stars  having a dispersion in metallicity and age centered on these mean values. 

The MS boxes have been limited in $(g - i)$ colour space by requiring:  
\begin{equation}
\begin{array}{rrcll}
0.20 & < & (g - i) & < 0.55. \\
\end{array}
\end{equation}
Here the blue boundary corresponds to the MSTO colour and the red boundary has been chosen to minimize contamination by Virgo 
globular clusters (whose position in the CMD is indicated in the right panel of Figure \ref{FigCMD}; see also Figure~1 of \citealt{durr14}). By moving 
this selection box vertically in the CMD, we can identify MS stars at a range of heliocentric distances. When
doing so, the vertical widths of the boxes are scaled smoothly as a function of magnitude to ensure that the physical depth of each sample of stars selected within the boxes is 5 kpc.  Examples of MS boxes at three different distances are shown in the right panel of Figure~\ref{FigCMD}. 

For stars selected inside any of the MS boxes, we impose
a second constraint: a colour selection of 
\begin{equation}
\begin{array}{rrcll}
0.58 & < & (u^{*} - g) & < 0.90 \\
\end{array}
\end{equation}
that eliminates  disk white dwarfs from our sample of MSTO halo stars. In Figure~\ref{FigColourColour}, this additional colour 
selection in shown in the $(u^{*} - g)$--$(g - i)$ colour-colour diagram.  The location of white
dwarf stars and background galaxies is shown, demonstrating that, although
the white dwarfs and MS stars have similar $(g - i)$ colours, they are
cleanly separated using the $(u^{*} - g)$ index. Especially at fainter magnitudes, we find
that up to $\sim$~20\% of our MS samples would actually be (disk) white dwarf candidates if
this extra color selection criterion were ignored. Such a contamination rate would heavily skew the distance estimates for these stars, 
as demonstrated in Figure~\ref{WDdistcomp}. In this figure, we calculate MSTO
distances to stars in the TRILEGAL model
mock catalog \citep{gira05,gira12}  from which we took a selection of stars
comparable in galactic latitude and magnitude range to our NGVS sample. The
distance is calculated from the difference between the apparent magnitude of
MSTO stars in the model (selected by $0.1 < (g - i) < 0.35$) and an absolute magnitude of $i=4$, 
in accordance with the approximate MSTO absolute magnitude of halo stars with
age $\sim$ 11~Gyrs and [Fe/H] $\sim$ -1.0.  In order to derive the needed accurate distances to individual stars - rather than a characterisation of the distance distribution of the population - we focus for the purpose of this test solely on the main-sequence turn-off stars and thus use a $(g - i)$ range smaller than that used for the MS boxes.
The two panels show the derived and model input
distances both with and without an extra $(u^{*} - g)$ cut applied to the
model stars. Clearly, for deep photometric studies of halo star populations, 
the addition of a second color index that includes a blue bandpass is absolutely essential. 

Background galaxies in our sample of MSTO halo stars are also reduced by the combined colour cuts, but extend into the MSTO region shown in Figure~\ref{FigColourColour} with increasing number at fainter magnitudes.  Significant contamination from these misidentified background galaxies occurs at $i\sim23.5$ (see Fig~\ref{FigCMD}), therefore we adopt this as the depth limit to our analysis.

The effectiveness of this approach in eliminating contamination from background sources, including globular
clusters in the Virgo cluster, is illustrated in Figure~\ref{FigGalaxies}.
In the left panel of this figure, we show the location of Virgo galaxies within the NGVS footprint,
with symbol sizes scaled according to galaxy $B$-band luminosity.  The supergiant elliptical galaxies M49 
and M87 are denoted by the red crosses in each panel of this figure.  The
middle panel shows a density map of the NGVS point sources brighter than $i$ = 23.5 mag, a selection that 
 eliminates most of the contamination from background galaxies.  However, it is clear that significant
contamination from misidentified Virgo globular clusters remains, manifesting as strong density 
peaks at the location of M49,  M87, and many other
Virgo galaxies. In the right panel of Figure \ref{FigGalaxies}, we show a map of MS stars selected 
using both a selection on $i$-band magnitude and on
location in the $(u^{*} - g)$--$(g - i)$ colour-colour diagram.
Clearly, the contamination from globular clusters has now been largely
eliminated by the colour-colour selection and no longer affects
our view of the intervening halo. 

In the remainder of the paper, we adopt the MS star selection process described above, which allows 
the halo to be traced out to distances of $d_{\odot} \sim$~90 kpc. Because NGVS is several magnitudes deeper than SDSS -- the 95\% completeness
limit of SDSS is reached at $i=21.3$ \citep{abaz09} -- it allows us to probe a
lot deeper into the outer halo. Main-sequence mapping based on SDSS photometry
alone has mainly focussed on distances $< 20$ kpc \citep[e.g.,][]{juri08}.

\subsection{Blue Horizontal Branch stars as tracers}
We examined the possibility of additionally using BHB stars as stellar
tracers. SDSS studies of MS stars have often been accompanied by BHB star studies. Due to their relatively high luminosities, 
BHB stars allow one to probe larger distances than is possible with MS stars. Unfortunately, the Megacam $u^{*}$
filter extends to redder wavelengths than the SDSS $u$ filter, compromising our ability to include BHB stars in our analysis: i.e., the 
additional red coverage includes a significant region redward of
the Balmer break, whereas in SDSS most of the $u$-band transmission is
dominated by light {\it blueward} of the break. This difference --- though
small in terms of wavelength coverage --- is important for BHB studies
because it means that the filter loses its gravity sensitivity
for these hot stars. We are therefore unable to separate
BHB stars from contaminating blue straggler stars (which have a similar temperature but a
much higher gravity) and therefore cannot use BHB stars as a complementary tracer population. 

\subsection{Halo Tomography}

In Figure \ref{FigMaps}, we use our selection criteria for MS boxes placed at
different distances to perform tomography of the Milky Way halo. Density maps for stars within each
distance slice are shown in 3\arcmin ~by 3\arcmin ~cells, smoothed with a Gaussian
filter of FWHM = 12\arcmin. The nine panels in 
Figure~\ref{FigMaps} show MS stars centered on our fiducial isochrone at mean heliocentric distances 
of $d_{\odot}$ = 5, 10, 15, 20, 25, 30, 40, 50 and 60~kpc. These distances are labelled in each panel, 
along with the mean MSTO magnitude in each bin.

Already at distances of 5--12 kpc,
an over-density begins to appear in the southern half of the
survey footprint, at a location and distance consistent with the VOD (see \S1).  At distances of 25--40 kpc,
the Sagittarius tidal stream is clearly seen slicing across the NGVS footprint.
At farther distances (up to 50 kpc), the Sagittarius tidal stream still appears at a low level. Although, a certain distance spread within the stream is seen, as well as expected (see also Sections \ref{sec:SagStreamDist} and \ref{sec:ComNumMod}), the tails of this distribution can be mostly ascribed to remnant contamination from white dwarf populations and photometric uncertainties.
At intermediate distances, we see a rather lumpy stellar distribution that could be due to a
mixture of stars belonging to these two substructures. We do not clearly detect the
bifurcation feature that runs parallel to the Sagittarius stream at slightly
higher declinations \citep[e.g.,][]{belo06,kopo12,slat13,debo15}. This is because, at the highest declinations probed 
by the NGVS, the secondary stream is too faint and no clear background region
is available within the NGVS footprint for it to stand out against. 

The VOD, which has previously been observed at distances ranging from 5 to 30 kpc, is known to spread over thousands of 
square degrees \citep{juri08,jerj13}. The VOD stellar density reaches its maximum at lower declinations 
than are accessible by the NGVS and, indeed, most  previous studies have targeted areas of declinations 
lower than those examined here.  We note that several formation models for the Sagittarius stream suggest that 
some trailing arm debris and/or older wraps could be present within the NGVS footprint at smaller distances 
than the leading arm. Based on photometry alone (i.e., without any radial velocity or proper motion information), it is difficult to 
conclusively rule out the possibility that some of the substructure we attribute to the VOD may instead be associated 
with the Sagittarius tidal debris. However, according to current models, any secondary streams are not expected 
to be as strong as the prominent over-density we see in our data. For instance, in the LM10 model, no debris 
features in the NGVS fields are predicted at $\sim$12-18 kpc, which is where we see our peak
density. Henceforth, we will therefore refer to the southerly feature detected between 5 and 25 kpc as the VOD
and consider it separately from the Sagittarius stream. 

We see the VOD first appear at $d_{\odot}\sim$ 8 kpc and disappear completely by $d_{\odot}\sim$ 25
kpc. In this distance range many smaller ``hotspot'' regions can be observed
within the broader over-density feature itself.  Spatially, the VOD is separated from the Sagittarius tidal stream
in distance as well as location on the sky. In the density maps shown in
Figure \ref{FigMaps}, the stream appears clearly in the mid-upper
region of the footprint, whereas the VOD is strongest at the very bottom of the
survey, at the declinations $\lesssim$ 8$^\circ$.
A sharp cut-off is used to indicate the strongest part of the VOD in {\it panel ii} of Figure~\ref{FigMaps}, though it appears across the entire region with density decreasing towards the top-right (see {\it panels ii} and {\it iii}).

As shown in, e.g., \citet{helm11}, galactic substructures tend to be distributed very
anisotropically on the sky in halo simulations carried out within a $\Lambda$CDM
framework. One should therefore be careful about na\"ively linking two
over-densities on the sky to a common physical
origin. Indeed, it may be prudent to look more deeply into regions that are
already known to show an abundance of substructures. Although this is
exactly what we have done in our analysis, it is worth noting that we
find no convincing evidence 
for new over-densities lying behind the Sagittarius leading arm.


\section{The Distance Gradient of the Sagittarius Tidal Stream}

In this section, we use our halo star density maps to examine the three-dimensional structure
of the Sagittarius tidal stream, specifically its distance gradient along the line of sight.

%
   \begin{figure*}
   \centering
   \includegraphics[scale=1.]{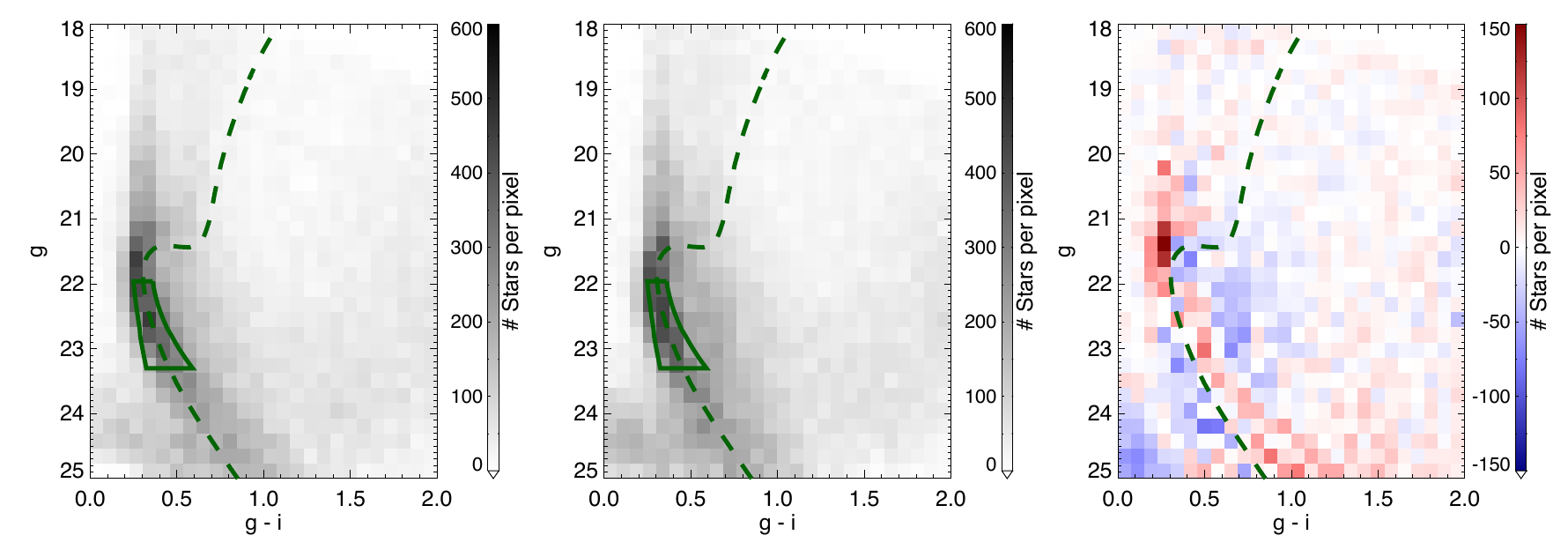}
   \caption{\emph{(Left and middle panels)} Colour-magnitude diagrams (CMDs) of the stars in two locations 
   		at either end of the Sagittarius tidal stream within the NGVS (see {\it panel vi} of Figure \ref{FigMaps}).  
		Background sources have been subtracted from the CMDs for clarity using two NGVS 
		reference fields (at declinations of $\delta \sim$ 21 and $\sim$ 27 deg). 
		A PARSEC isochrone (with an age of 9 Gyr and a metallicity of [M/H] = --0.7 dex; dashed green line) is plotted in both diagrams
	 	at a distance modulus of 17.8, along with the selection box used in
		the main sequence cross-correlation algorithm described in \S4.2.  The theoretical isochrone and the main-sequence selection box 
		have been placed at the location of the Sagittarius tidal stream, which has a clearly over-dense main sequence star population.
		\emph{(Right panel)} Differential CMD showing the residual of the CMDs in the previous panels.  Red  
		corresponds to a greater number stars in the first CMD, and blue corresponds to a greater number of stars in the 
		second CMD.  The colors have been scaled with the number of excess stars in either CMD following the two-toned color bar
		(with negative numbers referring to an excess of stars in the second CMD).  
		There is a clear vertical shift between the two populations around the main sequence of the theoretical CMD, indicating 
		a distance gradient to the Sagittarius tidal stream across the NGVS footprint.
		} 
         \label{FigSgrCMD}
   \end{figure*}
%

   \begin{figure*}
   \centering
   \includegraphics[width=4.3cm]{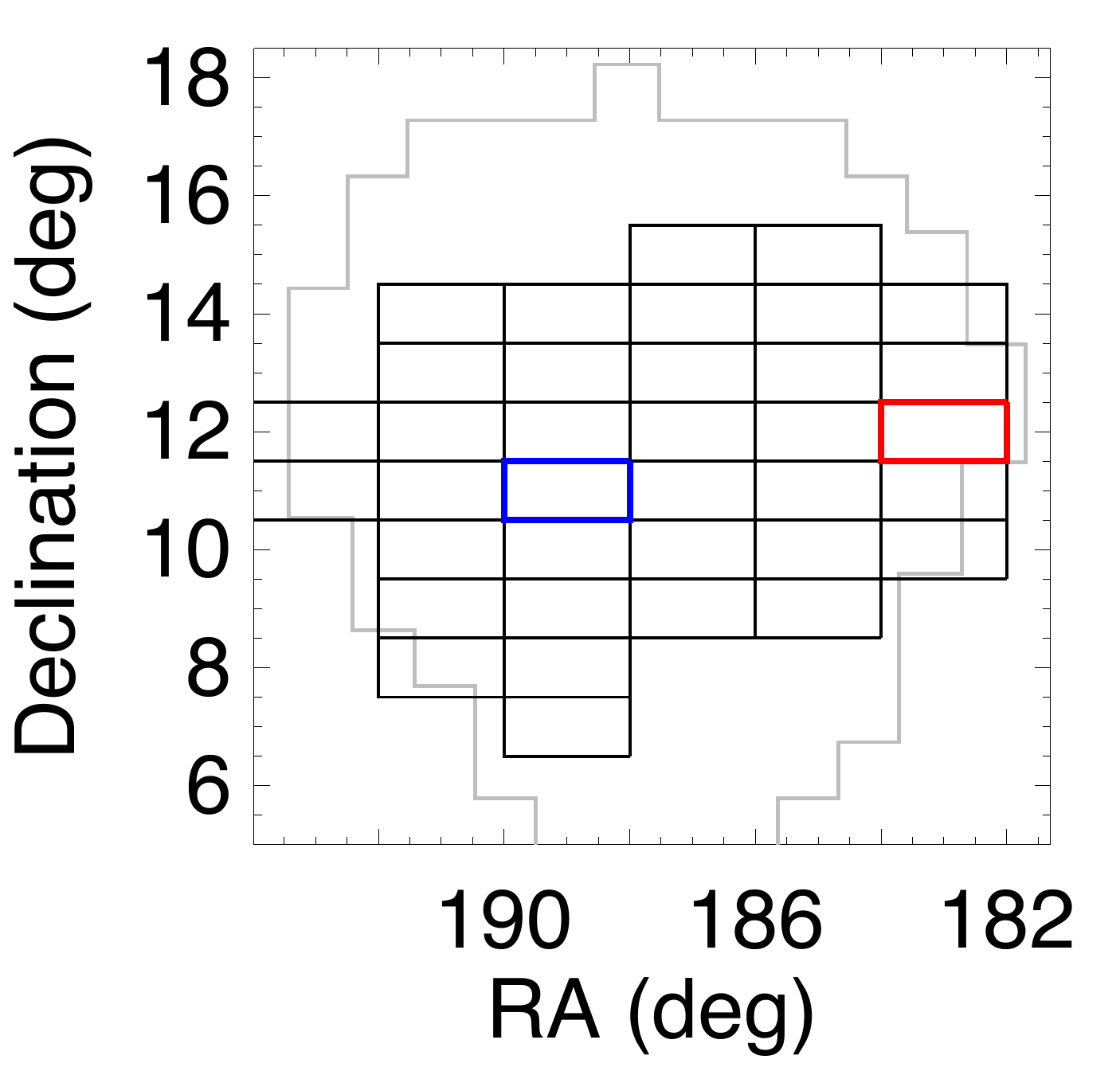}
      \includegraphics[width=12.5cm]{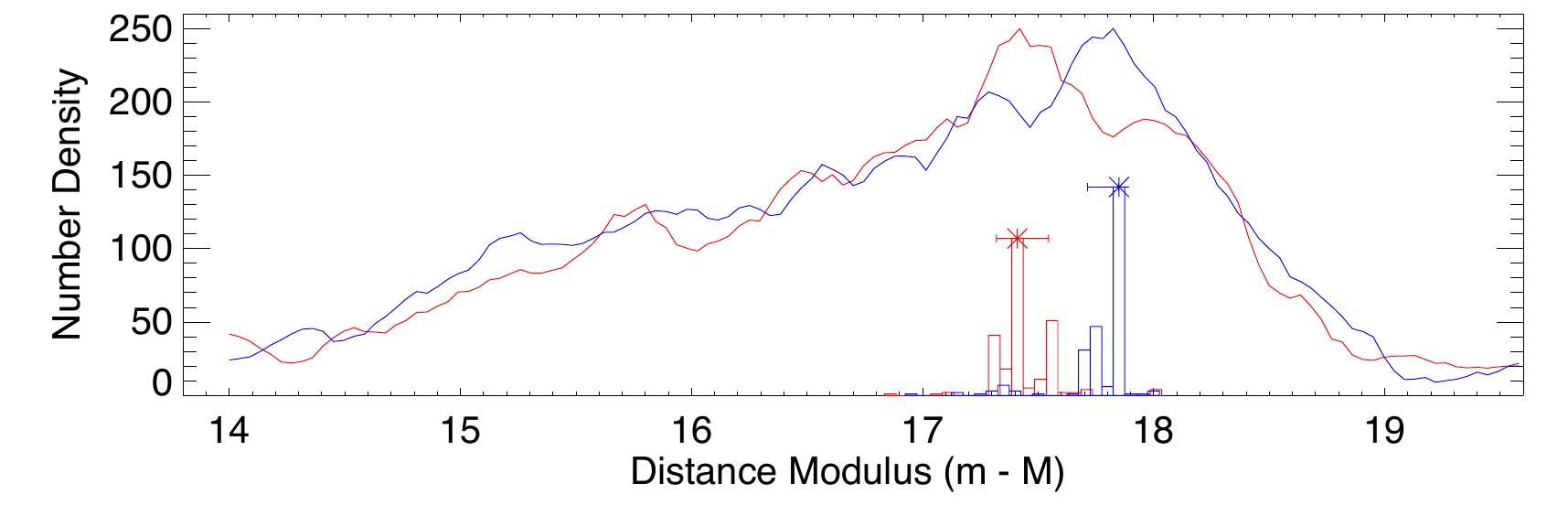}

   \caption{\emph{(Left panel)} Schematic diagram showing the 2 deg$^2$ regions used to tile the Sagittarius tidal stream,
   		each of which was used to calculate a mean stream distance.
   		\emph{(Right panel)} Number density plotted against distance modulus for Sagittarius tidal stream main sequence stars
		derived from stellar counts in the red and blue tiles shown in the preceding panel (red and blue curves, respectively).
		These luminosity functions were found via the main sequence cross-correlation algorithm outlined
		in Section 4.2.  Their maxima correspond to the distances of the tidal stream. This plot shows that 
		the Sagittarius tidal stream is located at a larger heliocentric distance in the blue (eastern) tile 
		than in the red (western) tile, a result consistent with the CMD analysis: see Figure \ref{FigSgrCMD}.
		The histograms result from a ``bootstrapping'' method to determine distance uncertainties: the cross-correlation 
		algorithm was performed on random 90\% completeness subsets (250 for each tile), yielding a distribution of peak distances
		from each sample.
		 The red (blue) histogram corresponds to the red (blue) luminosity function.
		} 
         \label{FigMSDC}
   \end{figure*}
%

\subsection{CMD Analysis of the Sagittarius Tidal Stream}\label{sec:CMDSgrdistgrad}

The Sagittarius stream first appears in Figure \ref{FigMaps} at a distance of $d_{\odot} \sim 20$~kpc
and reaches its greatest prominence in panel {\it vi}, at a distance of $\sim$ 30 kpc. The stream is clearly visible as a 
broad swath extending from  higher to lower declination as right ascension increases.  

Examination of the panels {\it vi-vii} of Figure \ref{FigMaps} show hints of
a significant distance gradient in the Sagittarius stream across the NGVS
field, with the largest right ascension region being systematically further
away than the other extreme. We explore this further in
Figure~\ref{FigSgrCMD}, where we show the CMDs of halo MS stars selected within the red and blue boxes 
drawn in panel {\it vi} of Figure~\ref{FigMaps}. The two NGVS
background fields at $\delta \ge 20^\circ$, which do not intersect any obvious substructures in this
distance range (see Figure \ref{FigBelNGVSFOS}), were used to create a reference CMD that was
subtracted from the CMDs of the tidal stream regions. These background-subtracted CMDs are shown
in the left  and middle panels of Figure~\ref{FigSgrCMD}. Our fiducial isochrone is shown as the dashed green
curve in each panel, shifted to a common distance modulus of $(m-M)$ = 17.8 mag ($d_{\odot}$ $\approx$ 36.3~kpc).

A differential CMD for the Sagittarius stream is shown in the right panel of Figure~\ref{FigSgrCMD}, in which the 
CMD in the middle panel (composed of stars within the blue spatial box in panel
{\it vi} of Figure~\ref{FigMaps}) was subtracted from the CMD in the left panel (composed of stars in the red
spatial box in panel {\it vi} of Figure~\ref{FigMaps}). The results of this
subtraction in the right panel of Figure~\ref{FigSgrCMD} are colour coded according to the differential over-
or underabundance of the number of stars in the red or blue spatial box respectively.
This differential analysis illustrates that the mean distance to the stream
changes with position on the sky. This is most clearly seen to the left and
right of the main sequence of the overplotted fiducial
isochrone. Whereas left of the isochrone main sequence the
pixels are mostly blue (corresponding to a relative overabundance of stars in the blue spatial
box at larger distance), they are mostly red on the right side, in line with a relative overabundance of stars in the red spatial
box at closer distances.  By fitting an isochrone to both spatial fields
separately, we find that the vertical shift between the two best fits
corresponds to a distance difference of $\sim$ 13 kpc.

\subsection{Main Sequence Turn-off Star Analysis}

\subsubsection{Refining our CMD analysis }

The deep photometry and excellent image quality available in the NGVS allow us to identify MS halo stars 
in this region of the sky --- with minimal contamination --- to larger
distances than has been possible in the past
\citep[e.g.,][]{juri08,sesa11}. Additionally, this excellent dataset enables
the measurement of distances to a higher level of precision than we have
demonstrated thus far using broad MS boxes.  We therefore present a more refined approach in this section.

We begin by fine tuning the box dimensions so that their width at each luminosity is proportional to the 
uncertainty of the photometric measurements (resulting in a trapezoid-like shape in the CMD; see the left
and middle panels of Figure~\ref{FigSgrCMD}).  This allows us to
compensate naturally for the larger photometric uncertainties of fainter stars when
assessing their likelihood as Sagittarius stream members. The aim of this exercise is to maximize our selection of
Sagittarius stream stars, so we use again our fiducial isochrone as the ridge line of this
MS box.  Subsequently, we step this spatial box down the CMD to represent populations at increasing distance, and at each step apply the weighting scheme
to effectively cross-correlate the observed stellar density on the CMD with that of the model isochrone.  At each step of the MS box, as the region shifts down
the CMD, the highest weights are assigned to stars lying along the ridge line of the
MS region (i.e., those most likely to be at the exact distance of the
MS region). The assigned weights decrease for stars that are more displaced
from the ridge line according to a gaussian function whereby the
width is set such that the edges correspond to a 3$\sigma$ level, following Equation 1 of \citet{pila14}. When implementing this algorithm, 
we apply the same selections on $(g-i)$ and $(u^* - g)$ color described in 
\S\ref{sec:explMWsubstr} (i.e., Equations 2 and 3).   At each position of the MS box, the integral of the weighted
stars inside the box represents the MS star density.

\subsubsection{Sagittarius Stream Distances}\label{sec:SagStreamDist}

In the above approach, a peak in the MS star density profile corresponds to an over-density at the distance 
modulus applied to the fiducial isochrone. To measure distances at various positions along the stream, 
we divide the NGVS footprint into 2 deg$^2$ cells and apply the algorithm to each cell separately. 
Figure~\ref{FigMSDC} illustrates
this process, emphasizing the distance gradient along the stream within
the NGVS footprint. In the left panel, two cells at either end of the
stream have been highlighted. In the right panel, we show  the corresponding MS
star density distributions for these regions. The two peaks are separated in
mean distance by 6 kpc. This is slightly smaller compared to the distance 
separation inferred from the CMD analysis described in Section
\ref{sec:CMDSgrdistgrad}, but note that the spatial boxes are different as well.

There are several sources of uncertainty intrinsic to this method. The first is the choice of theoretical 
isochrone used for the ridge line of the MS region \citep[which we have chosen specifically to match what is 
known about Sagittarius stream stars in this region of the sky, see e.g. ][]{Chou07}. Other sources of error 
include sampling errors caused by the limited number of stars available in a given box,
photometric errors on the individual stellar magnitudes and colors, and contamination by foreground and/or 
background sources. 

We have accounted for photometric uncertainties in our distance measurements by implementing 
a ``bootstrapping'' method. Random 90\% completeness stellar subsets
were selected and the cross-correlation algorithm was run on each of them to investigate the
robustness of the original distance estimate. This procedure has been repeated 250
times for each region. The red and blue histograms shown in the right panel
of Figure \ref{FigMSDC} are the 90\% completeness peak distance
distributions for the same two example regions used to create the
corresponding MS density distributions.  The percentiles of the 90\% completeness histograms are taken as the
uncertainties on our distance measurements throughout this work.

In cases where the bootstrapping method yielded more than one distinct
peak in MS star density distribution, the individual peaks were treated as separate
over-densities --- each with its own distance estimate and accompanying error
distribution. This approach acknowledges the possibility that any given line of sight
might cut through multiple over-dense regions, rather than forcing the
method to find only a single peak or, worse, defining a final
distance as the mean of two distinct peaks. However, we do define a
threshold above which a distance peak is considered to be a real feature. In what
follows, we consider only those over-densities that were identified by the bootstrapping method as a
main feature in more than 40\% of the cases. Because they should provide a useful test of future 
simulations, we have summarized our findings in Table 1. From left to right, the columns  of this table record 
the cell right ascension and declination, measured distance modulus,
upper and lower 1$\sigma$ errors, and adopted weights. These weights
correspond to the fraction of times the overdensity peak at this distance is recovered in our bootstrapping
tests.


%
   \begin{figure}
   \centering
    \includegraphics[width=9cm]{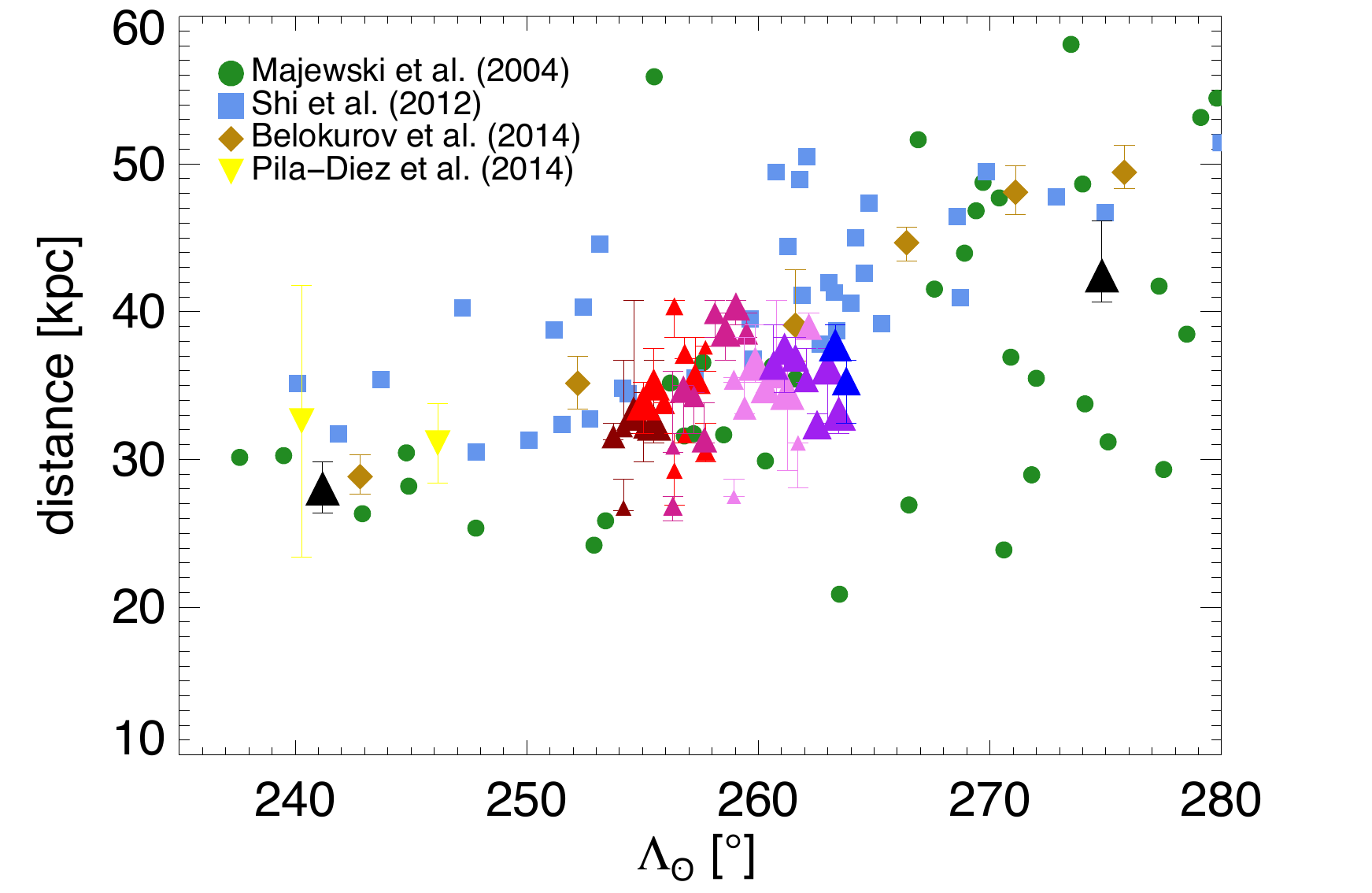}
       \caption{Comparison of distances to the Sagittarius tidal stream calculated in this work 
      		to those for 
		leading arm of the Sagittarius stream inferred from blue horizontal branch stars \citep{belo14}, 
		red horizontal branch stars \citep{shi12}, M giants \citep{maje04} and 
		main sequence near turn-off stars \citep{pila14}. Our measurements
		within the NGVS footprint are shown as colored triangles, while the two black 
		triangles show measurements made in the NGVS
		background fields. NGVS data points having the same right ascension
		are plotted using the same color.} 
         \label{FigSgrData}
   \end{figure}
%

%

\subsection{Comparison with Other Measurements}

A comparison of the distances derived in this work with those from
papers in the literature  is shown in Figure~\ref{FigSgrData}.
For convenience, the distances have been plotted against the Sagittarius stream
coordinate, $\Lambda_{\odot}$, which is a coordinate system defined along the
Sagittarius stream \citep[for details, see][]{maje03}.\footnote{Throughout this work, we
use the coordinate system defined by \citet{maje03}, rather than the
transformation given by \citet{belo14}.}  
Since only one of the Sagittarius stream coordinates is given in this figure on the
x-axis, a naive plotting of literature results will give a large range caused
by targets that are associated with leading or trailing arm debris
respectively (as well as older wraps). Here we show only stars associated with leading arm debris by the
various authors, as this is the component of the Sagittarius stream that we
see in the NGVS footprint. There is generally good
agreement among the different studies, especially when one considers that different tracers
 have been used in the various studies, each with their own typical distance uncertainties. Moreover, \citet{maje04} and \citet{shi12} use individual
stars as tracers, whereas our measurements, as well as the points shown for \citet{pila14}
and \citet{belo14}, measure distances to a population of tracer stars, which explains the
reduced dispersion in these latter measurements. The larger dispersion in \citet{maje04}
relative to the work of \citet{shi12} can further be explained by the fact that
\citet{shi12} measured only distances to stars that were spatially overlapping with
the LM10 model of the Sagittarius stream, whereas no such restriction was used
in the earlier work of \citet{maje04}.

\begin{figure}
   \centering
   \includegraphics[width=9cm]{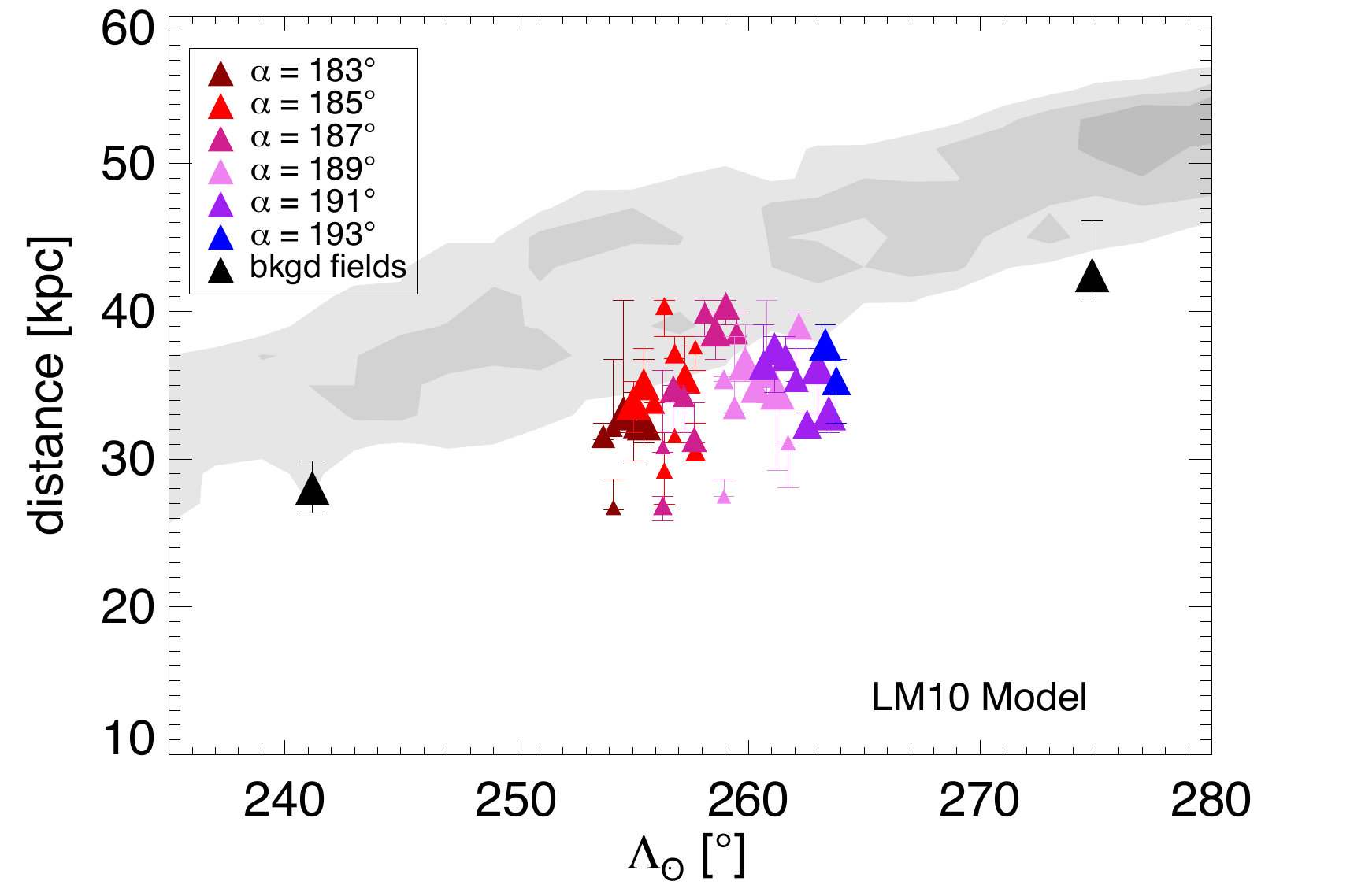}
   \includegraphics[width=9cm]{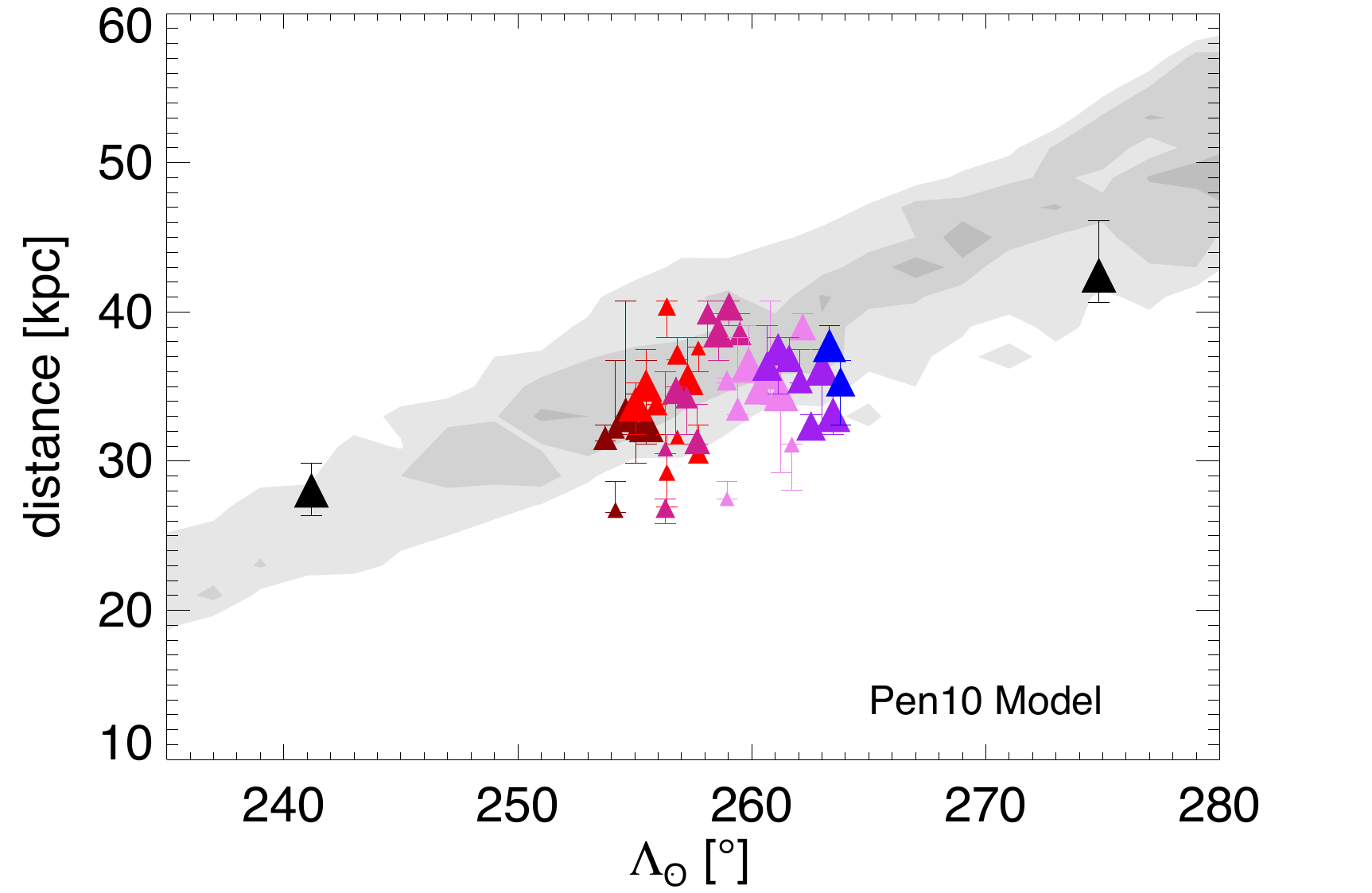}
   \includegraphics[width=9cm]{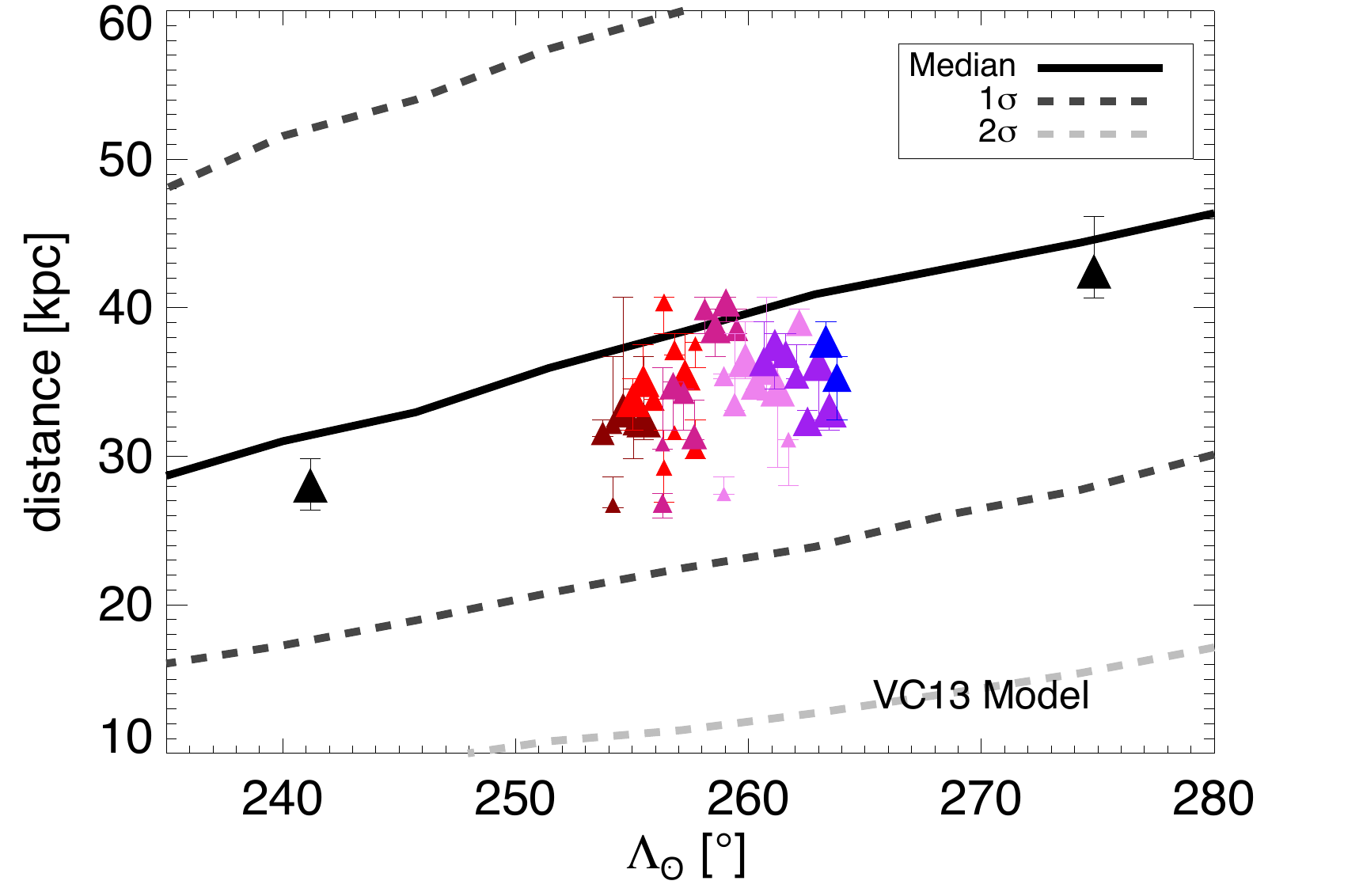}
      \caption{Distances to the Sagittarius tidal stream calculated in this work
        plotted aginst  $\Lambda_{\odot}$, the angular coordinate measured along the stream \citep{maje03}.
        Our measurements are plotted the same as in Figure~\ref{FigSgrData}. The three panels compare
        our measured distances to those from three different models of the Sagittarius tidal stream. Note that only
        distances with weights $>$ 40 have been plotted. 
		\emph{(Upper panel)} Particles from the N-body simulation of the Sagittarius tidal stream by (\citealt{law10} = LM10) are shown 
		as the gray-scaled density contours.  
		\emph{(Middle panel)}  Same as above, except using the simulation of (\citealt{pena10} = Pen10).
		\emph{(Lower panel)}  Median orbit (dark solid line) of the (\citet{vera13} = VC13) model for the Sagittarius tidal stream. 
		The 1$\sigma$  and 2$\sigma$ contours from this model are shown by the heavy and light dashed gray curves,
		respectively. 
		} 
         \label{FigSgrModels}
   \end{figure}

Our observed distances to the stream  are shown as the
colored triangles in Figure \ref{FigSgrData}, as well as in each panel of Figure \ref{FigSgrModels}. The symbol sizes
of each distance measurement have been scaled by the area under the corresponding peak in the bootstrapping method histogram.
To supplement measurements within the main body of the NGVS footprint, we also include the two NGVS
background fields situated along the Sagittarius stream (see Figure
\ref{FigBelNGVSFOS}). These are shown as black triangles offset by $\sim$ 15$^\circ$ from
the other NGVS data points.

\subsection{Comparison with Numerical Models}\label{sec:ComNumMod}

Figure~\ref{FigSgrModels} compares our measured distances to two
N-body models for the Sagittarius stream --- by \citet{law10} and \citet{pena10} --- as well as the Sagittarius
orbit model of \citet{vera13}.  The two N-body models, in the upper and middle panels,  are shown as
density contours (where the darker
gray corresponds to a higher density of model particles). In the lower
panel, the heavy black line corresponds to the median orbit of a test particle in the
potential considered by VC13. The 1$\sigma$ and 2$\sigma$ orbits from
the median are shown as the dashed dark and light grey lines. 

Our comparison between the model predictions and data points shows that the measured
distances to the Sagittarius stream are consistently smaller than
those predicted by the LM10 model. In fact, the densest part of the stream
predicted by the model does not significantly overlap with the data. The
Pen10 model --- which includes a rotation in the main body of the satellite --- provides a 
somewhat better match to the NGVS measurements, with significant
overlap between the densest part of the model and the data.  The
{\it gradient} in distance along the stream may be slightly shallower than
predicted by the Pen10 model, although this is mainly driven by the
two background fields and is within the boundaries of the N-body model. 
Both the LM10 and VC13 models, are systematically offset to further distances
than found in this analysis. These shifts are of order 10 kpc for the LM10 model and 5 kpc for the VC13 model (although here our data points do fit comfortably within the 
1$\sigma$ limits of the VC13 analysis). Such shifts are larger than what can be explained by photometric uncertainties or contamination of stellar populations alone. In fact, 
remnant contamination from white dwarfs in our analysis would shift our distance measurements in the opposite direction.

Unlike the potentials used by LM10 and Pen10, the potential of VC13
does not have its longest axis perpendicular to the disk of
the Milky Way. Rather, it is oblate and axisymmetric in its inner parts to ensure
the stability of the disk. The transition to a more triaxial outer halo is
smooth and --- including the influence of the LMC --- can have axis ratios
minor-to-major $>$~0.8 and intermediate-to-major $\sim$ 0.9, in good agreement
with findings from dark matter simulations. The agreement with our data
further strengthens that such models are worth exploring in more detail
with the benefit of detailed N-body simulations rather than a test-particle orbit. 


\begin{deluxetable*}{c c c c c c}
\tabletypesize{\scriptsize}
\tablecaption{Sagittarius Stream Distances Derived from the NGVS}
\label{table:1}      
\tablewidth{0pt}
\tablehead{
\colhead{$\alpha_{\rm center}$} & \colhead{$\delta_{\rm center}$} & \colhead{$(m - M)$} & \colhead{+$\sigma$} & \colhead{--$\sigma$} & \colhead{Weight (/100)}\\ 
\colhead{(deg)} & \colhead{(deg)} & \colhead{(mag)} & \colhead{(mag)} & \colhead{(mag)} &  
}
\startdata
     
     183.0 &      10.0 &     17.56 &      0.27 &      0.09 &       100 \\
     183.0 &      11.0 &     17.53 &      0.16 &      0.16 &        62 \\
     183.0 &      12.0 &     17.60 &      0.45 &      0.05 &       100 \\
     183.0 &      13.0 &     17.14 &      0.15 &      0.01 &        46 \\
     183.0 &      13.0 &     17.53 &      0.29 &      0.02 &        54 \\
     183.0 &      14.0 &     17.50 &      0.06 &      0.01 &        69 \\
     185.0 &       9.0 &     17.42 &      0.13 &      0.05 &        58 \\
     185.0 &       9.0 &     17.88 &      0.00 &      0.09 &        42 \\
     185.0 &      10.0 &     17.75 &      0.17 &      0.06 &        90 \\
     185.0 &      11.0 &     17.50 &      0.01 &      0.03 &        41 \\
     185.0 &      11.0 &     17.85 &      0.07 &      0.02 &        59 \\
     185.0 &      12.0 &     17.33 &      0.18 &      0.18 &        48 \\
     185.0 &      12.0 &     18.03 &      0.02 &      0.11 &        52 \\
     185.0 &      13.0 &     17.65 &      0.00 &      0.05 &        64 \\
     185.0 &      14.0 &     17.73 &      0.15 &      0.03 &        93 \\
     185.0 &      15.0 &     17.65 &      0.09 &      0.14 &       100 \\
     187.0 &       9.0 &     17.93 &      0.07 &      0.01 &        67 \\
     187.0 &      10.0 &     18.03 &      0.02 &      0.07 &        81 \\
     187.0 &      11.0 &     17.93 &      0.07 &      0.11 &        87 \\
     187.0 &      12.0 &     18.00 &      0.05 &      0.09 &        64 \\
     187.0 &      13.0 &     17.48 &      0.17 &      0.01 &        74 \\
     187.0 &      14.0 &     17.67 &      0.02 &      0.16 &        64 \\
     187.0 &      15.0 &     17.70 &      0.01 &      0.19 &        79 \\
     187.0 &      16.0 &     17.15 &      0.05 &      0.09 &        56 \\
     187.0 &      16.0 &     17.44 &      0.34 &      0.02 &        44 \\
     187.0 &      17.0 &     17.62 &      0.11 &      0.07 &        92 \\
     189.0 &       7.0 &     17.95 &      0.05 &      0.04 &        78 \\
     189.0 &       8.0 &     17.47 &      0.00 &      0.23 &        45 \\
     189.0 &       9.0 &     17.69 &      0.14 &      0.36 &       100 \\
     189.0 &      10.0 &     17.78 &      0.27 &      0.14 &       100 \\
     189.0 &      11.0 &     17.71 &      0.03 &      0.02 &        88 \\
     189.0 &      12.0 &     17.81 &      0.15 &      0.07 &        99 \\
     189.0 &      13.0 &     17.62 &      0.29 &      0.02 &        68 \\
     189.0 &      14.0 &     17.19 &      0.09 &      0.00 &        42 \\
     189.0 &      14.0 &     17.75 &      0.01 &      0.01 &        58 \\
     191.0 &       8.0 &     17.60 &      0.14 &      0.09 &       100 \\
     191.0 &       9.0 &     17.80 &      0.08 &      0.24 &        99 \\
     191.0 &      10.0 &     17.55 &      0.05 &      0.04 &        85 \\
     191.0 &      11.0 &     17.75 &      0.12 &      0.01 &        73 \\
     191.0 &      12.0 &     17.83 &      0.08 &      0.05 &        80 \\
     191.0 &      13.0 &     17.87 &      0.05 &      0.18 &        93 \\
     191.0 &      14.0 &     17.81 &      0.15 &      0.02 &        86 \\
     193.0 &      11.0 &     17.74 &      0.09 &      0.18 &        84 \\
     193.0 &      12.0 &     17.89 &      0.07 &      0.06 &        96 \\
     171.8\tablenotemark{a} &      20.8 &     17.24 &      0.14 &      0.14 &       100 \\
     203.4\tablenotemark{a}  &       6.8 &     18.14 &      0.18 &      0.09 &       100 \\

\enddata


\tablenotetext{a}{NGVS background fields.}

\end{deluxetable*}

\section{Summary and Conclusions}

We have used deep, multi-band ($u^*giz$) imaging from the Next Generation Virgo Cluster Survey (NGVS) 
to perform tomography of 
the Galactic halo in a $\sim$ 100 deg$^2$ region in the direction of the Virgo cluster. This 
region of the sky is known to contain two of the most prominent structures in the halo of
the Milky Way: the Virgo Over-Density (VOD) and the Sagittarius stellar stream. Compared to the 
SDSS, our NGVS photometry is both significantly deeper (by $\approx $ 3.2 mag for point sources) 
and has greatly improved image quality (0\farcs54 versus 1\farcs3). We describe an algorithm to select
candidate MSTO stars using a combination of source concentration, $i$-band magnitude
and $(g-i)$ and $(u^*-g)$ colors. Using this technique, we are able to identify MSTO halo 
stars out to distances of $d_{\odot} \sim 90$~kpc, with negligible contamination from  background sources such as
faint, compact galaxies and globular clusters in the Virgo cluster. 

We have carried out tomography of the halo using a MSTO star filtering approach. The VOD appears in 
our data at distances between 5 and 28 kpc and is notoriously clumpy in nature. It is most prominent 
at low declinations, but extends nearly to the highest declinations reached in the survey. By
contrast, the Sagittarius stream slices directly through the NGVS footprint at distances between 25 and 40 kpc.
Carefully dividing the stream into 2 deg$^2$ cells, we measure the distance gradient
along the leading arm and compare to the predictions of several numerical models. Our
distances are in somewhat better agreement with the model of \citet{pena10}
than that of \citet{law10}. Good agreement is also found with the
results of \citet{vera13} model, who provide a test-particle orbit of
the Sagittarius galaxy in a Milky Way potential that is of changing shape with radius and
includes the influence of the Large Magellenic Cloud. We find no evidence for new halo 
substructures beyond a distance of $d_{\odot} \sim$~40 kpc. 

Future papers in this series will explore the stellar kinematics in the NGVS and 
surrounding regions, as well as the properties of white dwarfs inside the NGVS footprint. 
For the time being, our tomographic analysis demonstrates how
deep, multi-band imaging at carefully chosen locations can be used to probe efficiently the
three dimensional geometry of halo streams. The
time is thus ripe for a newer generation of stream simulations that can be used to better
constrain the accretion history of the Milky Way and its gravitational potential. 

\acknowledgments
We would like to thank the anonymous referee for insightful comments that helped to improve this paper. ES acknowledges partial funding from the Canadian Institute for Advanced Research (CIFAR) Global Scholars Academy. 
The authors thank Carlos Vera-Ciro for very helpful discussions and for sharing the compilation data from \citet{vera13}.  ES and JFN thank the organizers of the KITP programme Galactic Archaeology and Precision Stellar Astrophysics for a stimulating workshop where part of this work has been completed. As such, this research was supported in part by the National Science Foundation under Grant No. NSF PHY11-25915.  CL acknowledges support from the National Natural Science Foundation of China (NSFC, Grant No. 11203017 and 11125313).
This work is supported in part by the French Agence Nationale de la Recherche (ANR) Grant Programme Blanc VIRAGE 
(ANR10-BLANC-0506-01), and by the Canadian Advanced Network for Astronomical Research (CANFAR), which has 
been made possible by funding from CANARIE under the Network-Enabled Platforms program. This research used the 
facilities of the Canadian Astronomy Data Center operated by the National Research Council of Canada with the support 
of the Canadian Space Agency. The NGVS team owes a debt of gratitude to the director and the staff of the 
Canada-France-Hawaii Telescope (CFHT), whose dedication, ingenuity, and expertise have helped make the survey a reality.

\textit{Facilities:} CFHT (MegaCam)
\\

\bibliographystyle{apj}
\bibliography{references2}

\clearpage

\end{document}